\documentclass[aps,twocolumn,longbibliography,superscriptaddress]{revtex4-2}

\usepackage{graphicx}
\usepackage[usenames,dvipsnames]{xcolor}
\usepackage{amssymb,amsfonts,amsmath}
\usepackage[colorlinks=true,citecolor=Cerulean,linkcolor=RubineRed,urlcolor=Cerulean]{hyperref}
\usepackage{epstopdf}
\usepackage{bm}
\usepackage{bbold}
\usepackage{upgreek}
\usepackage[inline]{enumitem}
\usepackage{booktabs}
\usepackage{multirow}
\usepackage[normalem]{ulem}
\usepackage{IEEEtrantools}

\usepackage{float}
\usepackage{verbatim}

\usepackage{lipsum, babel}
\usepackage{sidecap}

\usepackage{amsmath} 

\usepackage{hyperref}
\hypersetup{
    colorlinks=true,
    linkcolor=magenta,
    filecolor=magenta,      
    urlcolor=cyan,
    }

\usepackage{xr}
\makeatletter
\newcommand*{\addFileDependency}[1]{
  \typeout{(#1)}
  \@addtofilelist{#1}
  \IfFileExists{#1}{}{\typeout{No file #1.}}
}
\makeatother

\newcommand{\kBT}{k_\mathrm{B}T}
\newcommand{\kB}{k_\mathrm{B}}

\newcommand{\ket}[1]{\left|#1\right\rangle}

\newcommand{\ketbra}[2]{\left|#1\right\rangle \left\langle #2 \right|}

\newcommand{\EXP}[1]{~\times~10^{#1}}

\newlist{custominlist}{enumerate*}{1}
\setlist[custominlist]{label=(\roman*)}
\usepackage{filecontents}

\begin{document}
\title{Optical probing of phononic properties of a tin-vacancy color center in diamond}

\author{Cem Güney Torun}
\altaffiliation{These authors contributed equally to this work}
\affiliation{Department of Physics, Humboldt-Universit\"{a}t zu Berlin, 12489 Berlin, Germany}
\author{Joseph H. D. Munns}
\altaffiliation{These authors contributed equally to this work}
\affiliation{Department of Physics, Humboldt-Universit\"{a}t zu Berlin, 12489 Berlin, Germany}
\author{Franziska Marie Herrmann}
\affiliation{Department of Physics, Humboldt-Universit\"{a}t zu Berlin, 12489 Berlin, Germany}
\author{Viviana Villafane}
\affiliation{Walter Schottky Institute, School of Natural Sciences and MCQST, Technische Universität München, 85748 Garching, Germany}
\affiliation{Walter Schottky Institute, School of Computation, Information and Technology and MCQST, Technische Universität München, 85748 Garching, Germany
}
\author{Kai Müller}
\affiliation{Walter Schottky Institute, School of Computation, Information and Technology and MCQST, Technische Universität München, 85748 Garching, Germany
}
\author{Ulrich Kentsch}
\affiliation{Institute of Ion Beam Physics and Materials Research, Helmholtz-Zentrum Dresden-Rossendorf, D-01328 Dresden, Germany}
\author{Shavkat Akhmadaliev}
\affiliation{Institute of Ion Beam Physics and Materials Research, Helmholtz-Zentrum Dresden-Rossendorf, D-01328 Dresden, Germany}
\author{Anthony C. Withers}
\affiliation{Bayerisches Geoinstitut, University of Bayreuth, D-95440 Bayreuth, Germany}
\author{Andreas Thies}
\affiliation{Ferdinand-Braun-Institut (FBH), 12489 Berlin, Germany}
\author{Wentao Zhang}
\affiliation{Ferdinand-Braun-Institut (FBH), 12489 Berlin, Germany}
\author{Aleksei Tsarapkin}
\affiliation{Ferdinand-Braun-Institut (FBH), 12489 Berlin, Germany}
\author{Katja H\"{o}flich}
\affiliation{Ferdinand-Braun-Institut (FBH), 12489 Berlin, Germany}
\author{Tommaso Pregnolato}
\affiliation{Department of Physics, Humboldt-Universit\"{a}t zu Berlin, 12489 Berlin, Germany}
\affiliation{Ferdinand-Braun-Institut (FBH), 12489 Berlin, Germany}
\author{Gregor Pieplow}
\affiliation{Department of Physics, Humboldt-Universit\"{a}t zu Berlin, 12489 Berlin, Germany}
\author{Tim Schr\"{o}der}
\email[Corresponding author: ]{tim.schroeder@physik.hu-berlin.de}
\affiliation{Department of Physics, Humboldt-Universit\"{a}t zu Berlin, 12489 Berlin, Germany}
\affiliation{Ferdinand-Braun-Institut (FBH), 12489 Berlin, Germany}

\begin{abstract}    
        The coherence characteristics of a tin-vacancy color center (SnV) in diamond are investigated through optical means, including linewidth broadening effects and coherent population trapping (CPT) between the ground state orbital levels.
        Spectral analysis is required as due to the large spin-orbit splitting of the orbital ground states, thermalization between the ground states occurs at rates that are impractical to measure directly in the time domain.
        First, by implementing a temperature-dependent linewidth broadening measurement, including the challenging-to-measure D transition, phononic coupling coefficients are determined.  
        These measurements are performed on an emitter with a lifetime-limited linewidth and atom-like properties, making the measurement representative for high-quality SnVs.
        Next, a CPT-type experiment is carried out to independently analyze thermal decoherence processes at 4 K.
        The spectral information is transformed into its conjugate variable time, providing picosecond resolution and revealing an orbital depolarization timescale of ${\sim30{\rm~ps}}$.
        Consequences of the investigated dynamics are then used to estimate spin dephasing times limited by thermal effects.

\end{abstract}

\maketitle

    \begin{figure} 
            
            \centering
            \includegraphics[width=\linewidth]{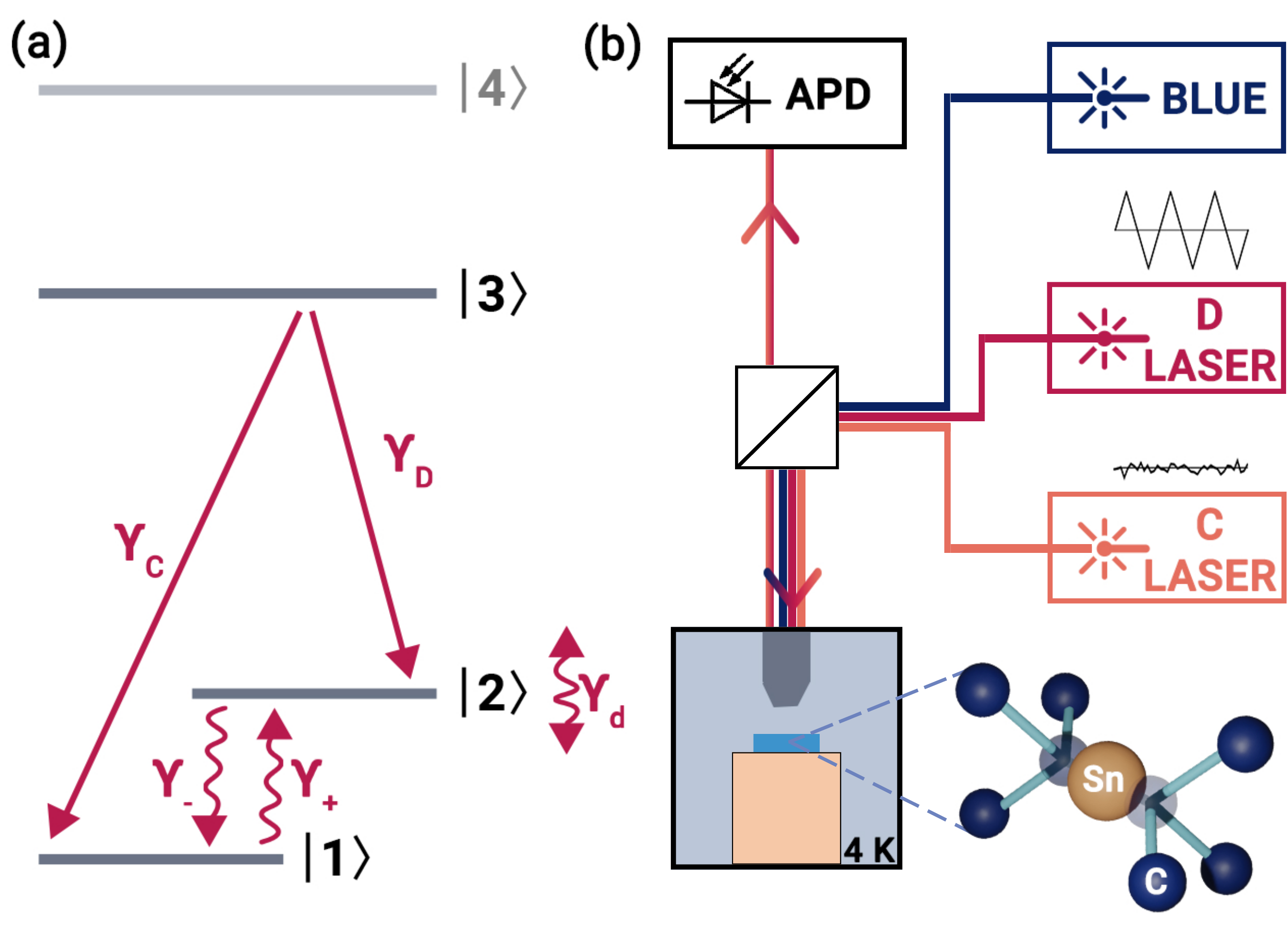}
            \caption{\label{fig:ExptScheme}  \protect\hypertarget{fig:ExptSchemeL}{} 
                (a) Orbital energy level structure and transitions of the SnV.
                $\gamma_{\rm C}$ and $\gamma_{\rm D}$ are the photonic relaxation (spontaneous emission) rates.
                $\gamma_{\rm +}$  and $\gamma_{\rm -}$ are phononic excitation and relaxation (depolarization) processes, respectively.
                $\gamma_{\rm d}$ is the dephasing driven by elastic two-phonon processes analogous to AC-Stark shift.
                (b) Simplified schematic for the experimental setup.
                Blue laser is kept on in the background to stabilize the charge environment and help with returning the resonance frequency to the set values after a hopping event occurs.
                C laser is tuned to the transition resonance and D laser scans the transition frequency in the coherent population trapping experiments. 
                The emitter is located either in a nanopillar or a solid immersion lens (SIL) that is cooled to 4~K.
                Collected fluorescence is measured with avalanche photodiodes (APDs).
            }
    \end{figure}

        \begin{figure*}
            \centering
            \includegraphics[width=\linewidth]{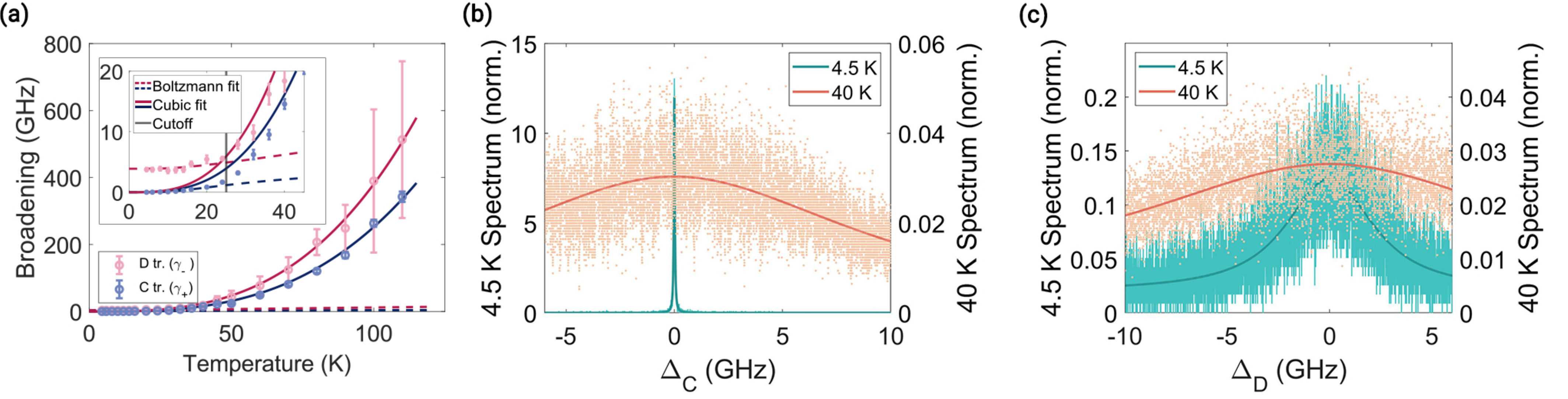}
            \caption{\label{fig:TDepCLine}
            \protect\hypertarget{fig:TDepCLineL}{}
               (a) Temperature dependence of the C and D transition broadenings extracted from optical linewidths (full width at half maximum).
                Measurements up to $T\leq40{\rm~K}$ are obtained from PLE measurements by scanning a resonant laser across the transition.
                At higher temperatures, the widths are obtained from PL spectra as the widths exceed the scan range.
                The error bars are extracted from the fit uncertainties.
                Different functions are used to fit the data while modeling single- and two-phonon regimes.
                INSET: At $T>25{\rm~K}$ (right side of the gray line), the data points deviate from the single-phonon model observed at lower temperatures, and the behavior is well-described by a cubic function.
                This suggests that two-phonon dephasing processes are dominant in this regime.
                (b), (c) Lineshapes of the C and the D transitions, respectively, at 4.5 and 40 K. Areas under the fitted curves are normalized to unity.
            }
        \end{figure*}

\section{Introduction}
    
    In the last decade group-IV color centers in diamond (G4Vs), especially the silicon-vacancy (SiV), have emerged as a promising candidate for quantum technologies \cite{SenkallaPRL2024, knaut_entanglement_2023}.
    The main advantage of this family of color centers stems from their inversion symmetric structure  within the diamond lattice \cite{Wahl.2020}, which lends increased resistance to environmental charge noise (spectral diffusion) compared to the more widely studied nitrogen-vacancy (NV) color centers  \cite{OrphalKobin2023PRX, Sipahigil2014PRL,Narita.2023}. This advantage becomes particularly salient for integrated applications \cite{Rugar2019PRB, Nguyen2019PRB, Martinez2022PRL, Bopp2024Adv}, where nanofabrication processes introduce surface charge noise in closer proximity to the color center.

    Although the G4Vs have exceptional optical qualities, the two branched orbital ground states of the G4V energy manifold (Fig.~\ref{fig:ExptScheme}\hyperlink{fig:ExptSchemeL}{(a)}) allow spin-conserving phononic transitions and result in a substantial decoherence channel for these quantum memories \cite{Jahnke2015NJP, Trusheim2020PRL}. In comparison, the NV is mainly protected from these thermal effects as the ground states have different spins and do not mix as quickly, enabling longer coherence times \cite{AbobeihNatComms2018, Karapatzakis2024PRX}.
    
    Among the G4Vs, the negatively charged tin-vacancy (SnV) center in diamond, the third heaviest in group IV, has demonstrated better coherence properties at temperatures around 1 K \cite{Guo2023PRX} compared to lighter SiVs \cite{Pingault2017NatCom} and germanium-vacancy centers \cite{Siyushev.2017}.
    The heavier dopant atom results in a large ground state splitting ${\sim850{\rm~GHz}}$ that is advantageous for the coherence properties of the spin states in the lower branch \cite{Jahnke2015NJP}.
    In particular, already at cryogenic temperatures of roughly 2 K, the SnV spin coherence times are not limited by phonon processes \cite{rosenthal_microwave_2023,Guo2023PRX,Debroux2021PRX}, in contrast with the mK temperatures required by SiVs \cite{Sukachev2017PRL, Becker2018PRL}.
    Furthermore, although the Debye-Waller factor is slightly reduced in comparison with the SiV (${\sim60}\%$ \cite{Goerlitz2020NJP} instead of ${\sim70}\%$ \cite{Neu2011NJP}), also as a consequence of the dopant size, higher quantum efficiency is expected due to reduced coupling to thermal processes \cite{Iwasaki2017PRL}.
    
    These properties pinpoint SnVs as a favorable platform for integrated photonic quantum information processing \cite{parker_diamond_2023, clark_nanoelectromechanical_2023,Li2024Nature}. 
    Understanding the decoherence dynamics of the SnV color centers, however, is imperative for developing more efficient techniques for coherent control.
    Therefore quantifying the phononic processes constitutes an important benchmark for the performance of G4V qubits.
    
    Directly measuring ultrafast events on the order of picoseconds, at which these phononic processes occur, in the time domain is challenging in terms of both control (pulse generation and synchronization) and photon detection jitter.
    A route to circumvent this is therefore to resolve these in the frequency domain instead, where transition rates show up in the form of resonance linewidths.

    In order to identify the underlying physical processes in the SnV, first the single-phonon and two-phonon linewidth broadening regimes are identified by monitoring the evolution of an optical transition linewidth as a function of temperature for the two observable optical transitions of the SnV. The so-called D transition has to date not been measured because it is particularly challenging to measure as the ground state of the transition has a very low thermal population at temperatures of a few kelvin. 
    
    Orbital depolarization rates are then measured through conducting coherent population trapping (CPT) \cite{SantoriPRL2006, Pingault2014PRL, Rogers2014PRL, Debroux2021PRX, GoerlitzNPJ2022, Wu2025} experiments. 
    Introducing CPT as an alternative method for measuring ultrafast decoherence processes to linewidth measurements via photoluminescence excitation (PLE) spectroscopy is particularly relevant in a variety of cases. For example, the population of one of the investigated levels can be too low for acquiring a signal during spectroscopy. Moreover, the emitters in the solid-state are usually in proximity to other defects that can ionize and actively modify the electrostatic environment \cite{Pieplow2022InPrep}. Resonant interaction on one of the transitions could create undesired environmental interactions due to a near-field enhancement \cite{Li2024NatPho}. Therefore, a less invasive approach is the detuned driving via CPT.
    
\section{Electron-phonon coupling} \label{sec:scope}
    The SnV is a four-level system under zero-magnetic field (Fig.~\ref{fig:ExptScheme}\hyperlink{fig:ExptSchemeL}{(a)}), similar to other G4Vs \cite{Thiering2018PRX}. It contains four orbital levels, two in the ground state and two in the excited state manifold. At cryogenic temperatures, only two photonic relaxation channels are observable: conventionally named C transition between levels $\ket{1}$ and $\ket{3}$ ($\gamma_{\rm C}$) and D transition between levels $\ket{2}$ and $\ket{3}$ ($\gamma_{\rm D}$).

    Decoherence mechanisms between the ground orbital levels of an SnV can be described in terms of relaxation between the two levels $\gamma_\pm$ and pure dephasing $\gamma_{\rm d}$ (Fig.~\ref{fig:ExptScheme}\hyperlink{fig:ExptSchemeL}{(a)}).
   For the orbital levels, these processes are dominantly mediated by interactions with $E$-symmetric mode phonons, which includes direct single-phonon and Raman-like two-phonon transitions between the orbitals, as well as closed two-phonon elastic processes \cite{Jahnke2015NJP}.
   Here, the first two result in transitions between the orbital levels and thus depolarization ($\gamma_\pm$), while the latter results in an AC-Stark type effect that can contribute to dephasing ($\gamma_{\rm d}$).

The relation between orbital transition rates $\gamma_+=\gamma_{-}\exp\left(-\frac{h\Delta_{12}}{\kBT}\right)$ is governed by the Boltzmann distribution \cite{Jahnke2015NJP}, where $h$ and $\kB$ are the Planck and Boltzmann constants, respectively, and $T$ represents the temperature.
The large ${\Delta_{12}=820}$~GHz ground state splitting of SnVs has therefore an inverse effect on these rates at cryogenic temperatures.
While a lower temperature results in an asymptotical convergence to a constant $\gamma_-$ value, it reduces the thermal occupation of phonons that have sufficient energy to excite between the levels ($\gamma_+$). 
As a result, thermal relaxation from the upper to the lower orbital ground state occurs on a much faster timescale.

Within the scope of this work, unless otherwise stated, all reported timescales and rates refer to orbital processes and not the spin.
However, since the phononic processes are spin-preserving, the measured values also have a direct correspondence to spin coherence times.
Dephasing between the spin levels occurs when a closed cycle of (spin-preserving) orbital transitions is complete and it is therefore possible to provide an upper bound for the spin dephasing time $T_{2,{\rm~spin}}^*$ by the orbital excitation time $T_{+}=1/\gamma_+$.

    \section{Experimental Details}
        The SnVs investigated in this work are selected from two separate samples. The first emitter E1 is located in a commercial CVD grown diamond that was implanted with 35 MeV Sn ions penetrating roughly 4.35 µm deep. The sample was then high-pressure-high-temperature treated at 7.9 GPa and 2100$^\circ$C.
        This treatment yielded a high density of optically stable emitters with near-lifetime-limited linewidths systematically showing less than 60\% inhomogeneous broadening ($<$50~MHz) compared to expected lifetime-limited linewidths. 
        E1 is embedded into a solid immersion lens \cite{Jamali2014}, and an optical power-broadening and phononic-component subtracted linewidth of 37.3(9)~MHz, matching well with the 32.0(8)~MHz lifetime-limited linewidth.
        This provides a spectroscopic indication that indistinguishable photons from a single emitter can be generated. 
        To quantify the probability of finding multiple spectrally overlapping emitters for networking purposes, an ensemble measurement from a ten times higher dose implanted area of the sample is conducted and a 7.5 (3.0) GHz (FWHM) wide distribution of emitter central frequencies is obtained.
        The second emitter E2 is from a standard vacuum annealed (1150 $^\circ$C) sample and is embedded into a 240 nm diameter nanopillar \cite{Babinec2010NatNT,RugarPRB2019}.
        
        Further details about the samples, characterization measurements, and nanostructure fabrication are provided in Appendix \ref{sec:ap_sample}.
        A simplified schematic of the experimental setup used in the CPT experiments is presented in Fig.~\ref{fig:ExptScheme}\hyperlink{fig:ExptSchemeL}{(b)} and further details are provided in Appendix \ref{sec:ap_setup}.

\section{Measurements}        
    \subsection{Temperature dependent spectroscopy} \label{sec:thermalBroadening}
        To gain insight into the interaction dynamics of an SnV with its phononic environment, spectroscopic measurements on emitter E1 are conducted.
        At temperatures 40~K and below, PLE spectra are recorded by repeatedly scanning the resonant laser over the emitter’s transition and counting the emitted photons via the phonon sideband. 
        The measurements at higher temperatures are acquired via the photoluminescence (PL) spectra collected under non-resonant excitation.
        Extracted phononic components from the linewidths are plotted in Fig.~\ref{fig:TDepCLine}\hyperlink{fig:TDepCLineL}{(a)}, and the broadening effect is illustrated in Fig.~\ref{fig:TDepCLine}\hyperlink{fig:TDepCLineL}{(b,c)}. Details of the data analysis are provided in Appendix~\ref{sec:ap_r2}.
        
        The acquired data for both transitions are fitted to two models for separating the single-phonon and two-phonon regimes using a cutoff temperature. The broadened linewidths by single-phonon interactions are fit to the following equations based on the Boltzmann distribution 
        \cite{Jahnke2015NJP, Wang2024PRL}:

        \begin{equation} \label{eq:boltzmann}
        \begin{aligned}
            \gamma_+=\alpha_+\omega_{12}^3[\exp(\hbar\omega_{12}/(k_{\rm B}T)-1)^{-1}] \\
            \gamma_-=\alpha_-\omega_{12}^3[\exp(\hbar\omega_{12}/(k_{\rm B}T)-1)^{-1}+1] 
        \end{aligned}
        \end{equation}
        where $\alpha_{\pm}$ is the electron-phonon coupling parameter in this regime, and $\omega_{12}=2\pi\Delta_{12}$. $\gamma_{+(-)}$ is defined to be the phononic broadening observed in the C (D) transition. The phononic interactions in the excited state manifold are negligible since the splitting is large: $\Delta_{34}\gg\Delta_{12}$.
        
        In order to determine the cutoff temperature, goodness of fit values ($R^2$) are monitored for the C transition while adding more data points towards higher temperatures, and the best value is observed at 24 K (see Appendix \ref{sec:ap_r2}). Note that the fit follows a curved trend between 14 and 24 K, making a linear approximation nonoptimal. The extracted $\alpha_-=(2\pi)^{-3} \times 7.0 (6) \times 10^{-9}$ GHz$^{-2}$ and $\alpha_+=(2\pi)^{-3} \times 3.8 (3) \times 10^{-9}$ GHz$^{-2}$ have a factor of $\sim$2 discrepancy, and the value $\alpha_-$ matches better with the previously reported value of $\alpha = (2\pi)^{-3} \times  7.51 \times 10^{-9}$ GHz$^{-2}$ \cite{Wang2024PRL}. Presently, we are not able to interpret the discrepancy, and further investigations are required. 
        
        The higher temperature data are fit using a cubic relation $\gamma_{+,-}=A_{+,-}\times T^3$ \cite{Jahnke2015NJP}, and $A_+=2.52 (8) \times 10^{-4}$ GHz/K$^3$, $A_-=3.8 (1) \times 10^{-4}$ GHz/K$^3$ are extracted. It can be observed that the fit values between 24 and 40 K diverge from the measured values, which can be attributed to a transition between single and two-phonon-dominant regimes. Using fit models with both the single-phonon equations from Eq.~\eqref{eq:boltzmann} and a cubic factor on the whole dataset yielded results where $\alpha_\pm$ converged to 0. This indicates that higher temperature data dominate the trend of the overall dataset.

\begin{figure} []
            \centering
            \includegraphics[width=0.8\linewidth]{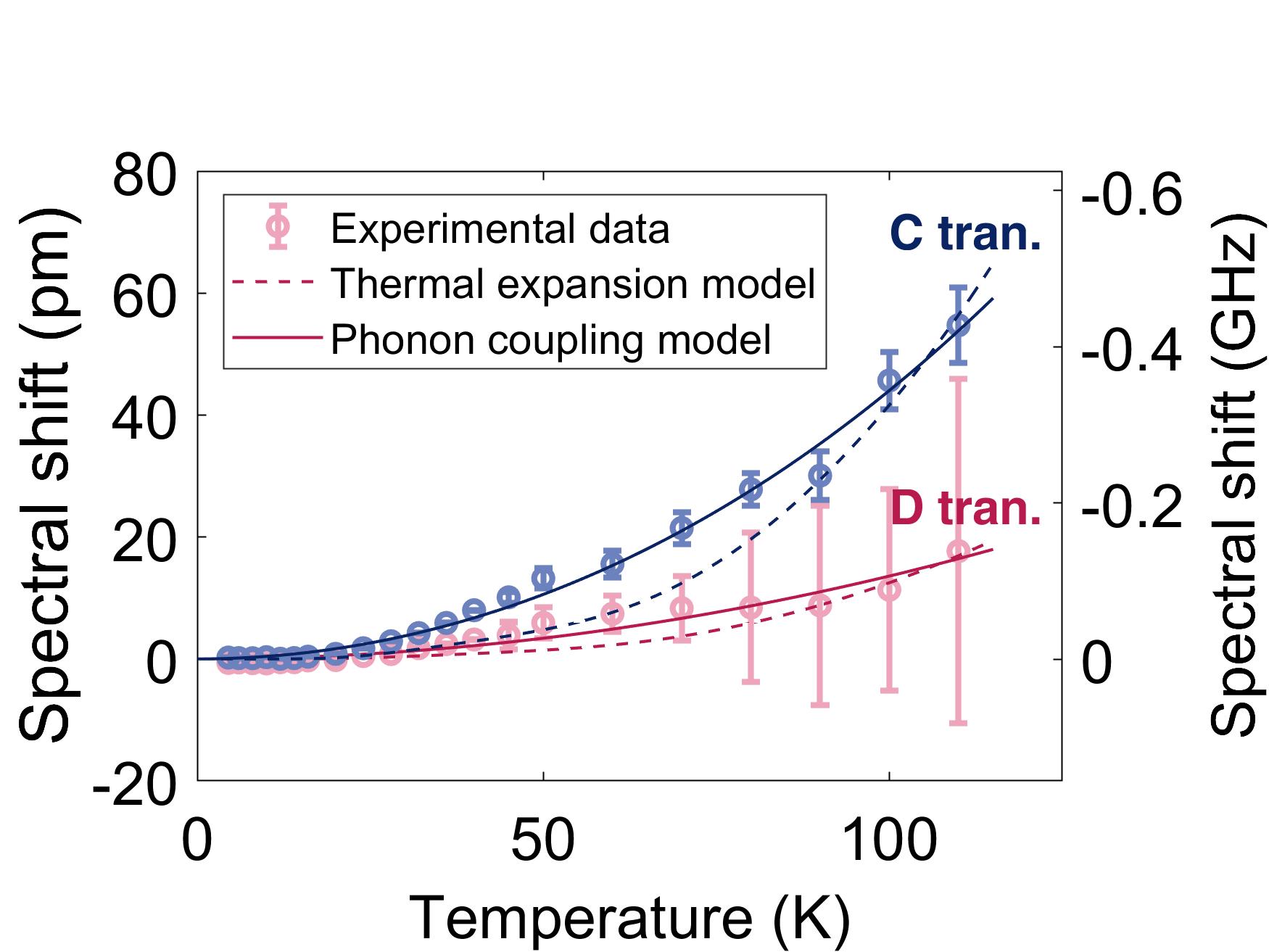}
            \caption{\label{fig:shift} 
            Temperature-dependent spectral shift of the C and D transition zero-phonon lines. The data are obtained from the same set of measurements from Fig.~\ref{fig:TDepCLine}. The central wavelength extrapolated to 0 K for the C (D) transition is 619.058 (620.111) nm. The error bars are extracted from the fit uncertainties. The data are fit to two models. The dashed lines correspond to a model of the diamond's thermal expansion. The solid lines correspond to the model for the SnV coupling to phonons using a polynomial function with quadratic and quartic contributions.}
            
    \end{figure}

    Temperature variations on optical transitions change not only the linewidth of the emission, but also the central wavelength.
    From the same spectroscopic data, we further analyze the spectral shift of both transitions as the temperature changes (Fig~\ref{fig:shift}). 
    The data are fit using two models \cite{Jahnke2015NJP}: (i) thermal expansion, and (ii) phononic interactions. 
    
    Model (i) uses the following functional dependence: $\Delta_{\rm E}= - A_{\rm pres} \times B_{\rm dia} \times E(T)$, where $\Delta_{\rm E} $ is the transition energy difference, $A_{\rm pres}$ is the hydrostatic pressure shift of the transition energy, $B_{\rm dia}$ is the bulk modulus of the diamond and $E(T)$ is temperature dependent expansion coefficient.
    The details of the analysis are provided in Appendix \ref{sec:ap_shift} and the fit from the thermal expansion model is shown with dashed lines in Fig. \ref{fig:shift}.
    It can be observed that thermal expansion does not successfully explain the observed trends, similar to previous measurements with the SiV \cite{Jahnke2015NJP}.

    Model (ii) represents the interaction of the SnV with the phonons using
    the following polynomial function (solid lines in Fig. \ref{fig:shift}): $A_{\rm C,D}\times T^2+B_{\rm C,D}\times T^4+\lambda_0$, where $A_{\rm C,D}$ represents the shifts by second-order linear shifts induced by $E$-symmetric phonons, $B_{\rm C,D}$ the shift induced by the $A_1$-symmetric phonons, and $\lambda_{\rm C,D}$ the central wavelengths at 0 K. The extracted coefficients are $A_{\rm C}=4.2 (7)\times10^{-6}$ nm/K$^2$, $B_{\rm C}=2 (6)\times10^{-11}$ nm/K$^4$, $A_{\rm D}=1.4 (6)\times10^{-6}$ nm/K$^2$, and $B_{\rm D}=0 (6)\times10^{-11}$ nm/K$^4$. 
    The phononic interaction model agrees better with the measured data compared to the expansion model.
    We attribute the dominance of the $A_{\rm C,D}$ term to the prevalence of the $E$-symmetric phonons below 110 K, as higher order polynomial terms have been shown to be more relevant at higher temperatures \cite{Goerlitz2020NJP,Alkahtani2018APL}.

        These results provide a detailed characterization of the temperature-dependent broadening and wavelength shift of the C transition and the previously spectroscopically inaccessible D transition in SnV centers. By combining PLE and PL spectroscopy, a temperature cutoff at 24 K is identified, indicating the transition from the single to the two-phonon regime in SnVs.
        Furthermore, the shifts in the optical transition wavelengths at the investigated temperature regime can be explained by the SnV coupling to $E$-symmetric phonons.

        \begin{figure*}[t]
            \centering
            \includegraphics[width=\textwidth]{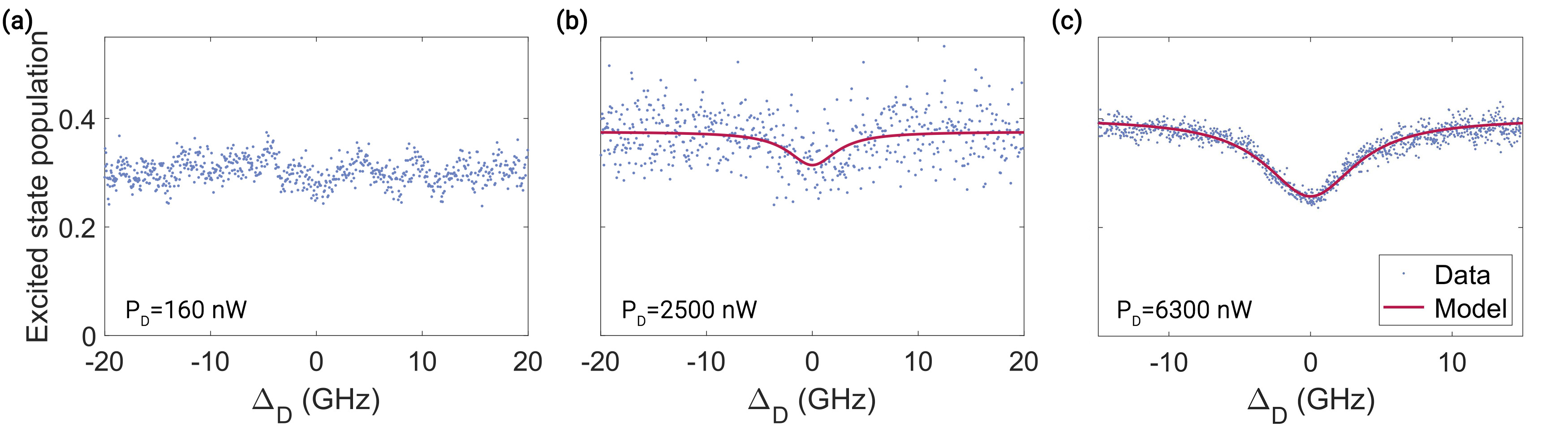}
            \caption{\label{fig:CPTDips}
            \protect\hypertarget{fig:CPTDipsL}{}
                Orbital CPT measurements at 4 K and varying scanning laser power at D transition. 
                The solid line is based on a fit from a Lindbladian master equation solution. 
                Increasing the power results with a higher Rabi frequency on the D transition and allows the formation of a coherent dip against the rapid orbital decay process. 
                Dataset indices from Table \ref{tab:}: (a) 13, (b) 9, and (c) 17.
            }
        \end{figure*}

    \subsection{Orbital CPT for ultrafast phononic relaxation rate estimation \label{sec:OrbitalCPT}}

        Measurement of relaxation times on the order of picoseconds is typically done with pump-probe measurements on ensembles of emitters. Accessing single emitter dynamics is therefore not feasible with established photoluminescence experiments. To overcome this limit, we introduce here a measurement in the frequency domain based on CPT on the orbital states to access the time domain. Using CPT to access spectral information has several advantages compared to using PLE, including probing non-populated levels  and detuned driving.
        
        The CPT effect occurs when two transitions with a shared excited state are driven simultaneously in a lambda energy configuration (e.g. between levels 1, 2 and 3, coupled by transitions C and D, Fig.~\ref{fig:ExptScheme}\hyperlink{fig:ExptSchemeL}{(a)}). 
        As one of the driving lasers is kept on resonance, a reduction in the fluorescence signal is observed when a second laser scans across the other transition frequency.
        Such a scheme results in bright and dark eigenstates, where only the bright state can be optically coupled to the excited state and produces a fluorescence signal.
        How much of the population becomes optically trapped in the dark state is determined by the coherence of these levels, and this directly influences the properties of the observed dip.
        
        In our measurements, the C laser is kept resonant with the C transition, while the D laser is scanned on the D transition.
        Blue charge state re-initialization light at 450~nm is applied throughout the scan \cite{GoerlitzNPJ2022}.
        The sample is cooled down to 4 K (see Fig.~\ref{fig:ExptScheme}\hyperlink{fig:ExptSchemeL}{(b)}).
        These measurements are performed twice (i.e. scanned in both directions) on emitter E2, using nine different power configurations.
        The data are fit by numerically integrating a Lindblad master equation until the steady state is obtained (details of the model are provided in Appendix \ref{sec:ap_model}).
        
        A direct correspondence between the powers and Rabi oscillations could not be used due to experimental instabilities (see Appendix \ref{sec:ap_setup} for details).
        Therefore, these fits include three parameters: 
        \begin{custominlist}
            \item Rabi frequency of the C transition $\Omega_{\rm C}$; 
            \item Rabi frequency of the D transition $\Omega_{\rm D}$; and 
            \item orbital phononic relaxation rate $\gamma_{-}$.
        \end{custominlist}
        A qualitative discussion of factors affecting the CPT profile is given in Appendix \ref{sec:ap_parameters}. There, it is shown that  all three parameters have distinct effects on the CPT dip: $\Omega_{\rm C}$ determines the offset of the CPT tail value, $\Omega_{\rm D}$ affects both the linewidth and the contrast of the dip, and  $\gamma_{-}$ only affects the dip contrast.  Distinct effects of the parameters enable a multidimensional fitting algorithm.
        
        Here, $\gamma_{+}$ is related to $\gamma_-$ by the thermalization rate determined by the Boltzmann distribution, and $\gamma_{\rm d}$ is set to 0.
        The latter constraint can be applied because at this temperature below 24 K single phononic processes are expected to be dominant, as reflected in the linewidth measurements reported in the previous section.

        Three example measurements and their fits are presented in Fig.~\ref{fig:CPTDips}.
        The steps of the data analysis and remaining measurements are presented in Appendix \ref{sec:ap_dataAnalysis} and \ref{sec:ap_allCPT}, respectively, and the fit results are provided in Appendix \ref{app:fitResults}.
        The quoted powers are the powers incident at the cryostat window, measured at the beginning of each experiment.
        Out of 18 measurements, 16 measurements are taken into consideration for the estimation of $\gamma_{\pm}$, as in the limit of low excitation power (e.g. Fig.~\ref{fig:CPTDips}\hyperlink{fig:CPTDipsL}{(a)}) the signal is considered too small. 
        On average (standard deviation), an orbital excitation time ${T_{+}=958}$~(138)~ns and orbital relaxation time ${T_{-}=31}$~(5)~ps are found, showing that the phononic relaxation is indeed an ultrafast process.

\section{Conclusion}

    In this article, decoherence dynamics based on phononic interactions of SnV color centers are investigated.
    Temperature dependent linewidth broadening for an SnV center is measured to determine the single and two-phonon interaction regimes, including the challenging to measure D transition. 
    
    We identify that single phonon processes are dominant at 4~K, and that a two-phonon cubic scaling starts dominating around 24~K. Through these measurements we also experimentally quantify, for the first time with an SnV, the phonon coupling rates at 4~K, the usual operating temperature for typical helium-cryostats. Consequently, the observed time scales are relevant for many applications. Using Eq.~\ref{eq:boltzmann} and extracted $\alpha_+$ enables further estimating the timescale for $T_+$ at 1.8 K, another standard operating temperature for cryostats, to 91 (8) ms. Since this value is around an order of magnitude higher than the currently highest recorded spin coherence time of 10 ms for SnVs \cite{Karapatzakis2024PRX}, it can be stated that the phonons will not be a limiting factor at 1.8 K or below \cite{rosenthal_microwave_2023}.
    
    CPT experiments are presented to show coherence between orbital levels and quantify the orbital phononic depolarization timescales.
    In particular, the orbital relaxation ($\gamma_{\rm -}$), which occurs in a few tens of picoseconds, is indirectly measured in the frequency domain, as it is very demanding to probe such rapid events in the time domain.  
    A time scale of 30~ps is ten times shorter than the electronic jitter of standard avalanche photodiodes and at the resolution limit of state-of-the-art superconducting nanowire detectors. We therefore confirm the improved resolution possible with CPT, a promising advantage of CPT even to time-resolved measurements. We further verify the estimated rates by measuring the broadening induced by phononic processes to the D transition linewidth and extract a consistent result in Appendix \ref{sec: ap_Dbroadening}. 

    Our measurements directly indicate that the upper branch of the SnV's orbital ground states is relatively short lived and therefore does not facilitate a practical qubit, similarly to previous observations with SiVs \cite{Becker2016NatCom}. The increased ground state splitting limits $T_{1}$ to the ${T_{-}\sim30}$~ps characterized in this work which is shorter than the coherence times of spin qubits.

    The extracted $T_{+}$ value of 958~ns also yields insight into achievable $T_{2,\rm spin}^\ast$ time scales as orbital excitation would provide an upper limit to the spin coherence times, when phonon-mediated dephasing is the limiting factor and other sources of dephasing (e.g. surrounding spin-bath \cite{Becker2018PRL} or gate-errors \cite{Debroux2021PRX}) play a minor role. This relation shows that CPT at zero magnetic field is a valuable method to estimate spin qubit decoherence, without directly accessing the levels themselves.

\vspace{\baselineskip} 
\section*{Data availability} The data that support the findings of this article are openly available at \cite{Zenodo2026}.

\begin{acknowledgements}
    The authors would like to thank Mariano Isaza Monsalve for insightful discussions regarding CPT dynamics; Laura Orphal-Kobin for sharing her experience as the PLE scans were implemented; Kilian Unterguggenberger with discussions regarding linewidth data analyses; Maarten van der Hoeven for his help setting the laser infrastructure; Mathias Matalla, Natalia Kemf, Nico Sluka, and Karina Ickert for support with lithography; Ina Ostermay, Adrian Runge and Dominik Sudau for performing the SiN$_{\rm x}$ deposition; Alex Kühlberg for performing the ion implantation; Ralph-Stephan Unger, Kevin Kunkel and Natalia Sabelfeld for their support with the plasma etching; and Sarah Benbouabdellah for helping with setting up the temperature scanning sample mount.

    The authors acknowledge funding by the European Research Council (ERC, Starting Grant project QUREP, No. 851810), the German Federal Ministry of Education and Research (BMBF, project DiNOQuant, No. 13N14921; project QPIS, No. 16KISQ032K; project QPIC-1, No. 13N15858), and the Einstein Foundation Berlin (Einstein Research Unit on Quantum Devices).
\end{acknowledgements}

\appendix

\section{Materials and Experimental Methods}

\subsection{Samples} \label{sec:ap_sample}

\begin{figure}
    \centering
    \includegraphics[width=0.33\textwidth]{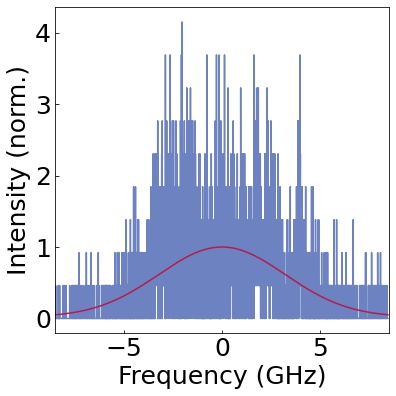}
    \caption{\label{fig:inhom_dist} 
        Inhomogeneous distribution of central frequencies from sample E2x2\_2\_A. Full-width-half-maximum of 7.5 (3.0) GHz is measured at a higher dose implanted part (10$^{10}$ ions/cm$^{2}$) of the sample.
    }
\end{figure}

The first sample used in this work for the linewidth broadening experiments, housing emitter E1 is named E2x2\_2\_A.
The SnV color centers are fabricated in a IIa-type (100) single-crystal electronic grade diamond. The sample surface is initially etched in Cl$_{2}$- and O$_{2}$-based plasma in order to remove any internal strain caused by the mechanical polishing \cite{Heupel2022pss}. The $^{120}$Sn ions are then implanted into the diamond with an energy of 35 MeV, corresponding to a final depth of about 4.35 µm as simulated by Stopping and Range of Ions in Matter \cite{Ziegler2010NIM} (SRIM) simulations, and with nominal fluence of 10$^{9}$ ions/cm$^{2}$. The sample is then annealed at 2100 °C under a pressure of 7.9 GPa for about 20 min, followed by a cleaning step in a boiling triacid solution of H$_{2}$S$_{4}$:HNO$_{3}$:HClO$_{4}$ (1:1:1) in order to remove possible graphitic layers from the surface that are formed during the annealing step. Solid immersion lenses (SILs) are finally fabricated by focused ion beam (FIB) milling \cite{Tsarapkin2026AFM}. The SILs have a hemispherical shape and a radius of 4.35 µm. In Fig.~\ref{fig:inhom_dist}, the distribution of the central frequencies from an ensemble of SnVs embedded in this sample is provided. The measurement is taken at a higher dose implanted part of the sample (10$^{10}$ ions/cm$^{2}$) and from a single diffraction-limited spot.

The second sample used in this work for the CPT experiments housing emitter E2 is an electronic grade, single-crystal diamond grown by chemical vapor deposition with \{100\} faces. The substrate is initially cleaned for about one hour in a boiling triacid solution (H$_2$SO$_4$:HNO$_3$:HClO$_4$, 1:1:1) \cite{Brown2019} and then etched in Cl$_2$/He and O$_2$/CF$_4$ plasmas, in order to remove any organic contaminants and structural defects from the surface \cite{Atikian2014}. Tin ions are subsequently implanted in the diamond, by employing the following nominal parameters: fluence of ${5\EXP{10}}$ atoms${\rm~cm^{2}}$ and implantation energy of $400{\rm~keV}$. SRIM simulations estimate a penetration depth of about 100~nm with this energy. Finally, an annealing step is performed at a temperature $T=1050~^\circ$C for about 12 hours and at a pressure $P<7.5\EXP{-8}$~mbar.
Nanopillars are fabricated by a combination of e-beam lithography and plasma etching. First, 200~nm of SiN$_x$ are deposited on the surface of the diamond in an inductively coupled-plasma enhanced chemical vapor deposition system. After spin-coating the sample with 300~nm of electro-sensitive resist (ZEP520A), we expose pillars with nominal diameters ranging from 180~nm to 340~nm, in steps of 40~nm. After development, the pattern is transferred into the SiN$_x$ layer by a reactive ion etching plasma (10 sccm CF$_4$, RF power = 100~W, P = 1~Pa) and then etched into the diamond during an ICP process in O$_2$ plasma (80~sccm, ICP power = 750~W, RF power = 200 W, P = 0.3~Pa). The remaining nitride layer is finally removed in a solution of buffered HF.

    \subsection{Optical setup} \label{sec:ap_setup}
        
        Resonant excitation lights (at 619~nm) are generated from a continuous wave dye laser (Sirah Matisse, DCM in EPH/EG solution), home-built second harmonic generation (SHG) source converted from a $1238{\rm~nm}$ diode (Innolume dfb-1238-pm-40-OI, Covesion MSHG-1230-0.5-4) and a commercial SHG based laser (Toptica DL-SHG-Pro).
        The home-built SHG source provides a simple and cost-effective alternative to dye or optical parametric oscillator based sources at 619~nm, in a wavelength region where no commercial diode lasers exist.
        Dye laser (D laser) and home-built SHG laser (C laser) beams are overlapped orthogonally in polarization on a beamsplitter and fiber coupled in the CPT experiments.
        In the overlap setup, the power of the resonant lasers can be controlled with the half-wave plates.
        The leak-through is then monitored on a wavemeter (High Finesse WS7).
        In addition to the resonant lasers, a re-initialization blue laser at 450~nm (Thorlabs LP450-SF15 or Hübner Cobolt 06-MLD) and non-resonant excitation green laser at 520~nm  (DLnSec) are combined into the same mode as the resonant excitation in a fiber wavelength combiner. For the temperature dependent linewidth broadening experiments, different laser colors are spatially combined on dichroic mirrors.
        Experiments are controlled via the Qudi \cite{Binder2017SoftwareX} software suite.
        
        The excitation mode is directed to the sample in a closed-cycle helium cryostat (Montana s50) equipped with an agile temperature mount via a home-built confocal scanning microscopy setup with an objective (Zeiss) of 0.9 numerical aperture.
        A quarter- and half-wave plate combination is employed to optimize the laser polarizations with respect to the SnV transition polarizations, where the C and D dipoles are orthogonal for a (100) terminated sample during the CPT experiments.
    
        For the CPT and PLE measurements a long-pass filter at 635~nm is used to filter the laser light, and the phonon sideband is coupled to the detection fiber.
        In PL measurements, a long-pass filter at 600~nm is used instead.
        In the detection fiber, the signal is split through a 50:50 fiber beam-splitter, where one port is used to monitor the signal counts (Excelitas SPCM-AQ4C), and the other port is directed either to a second channel on the APD for autocorrelation measurements, or to a spectrometer (Andor SR-500i-C-SIL and iDus camera) for PL spectrum measurements.

    \paragraph*{System instabilities:}
        On different CPT measurements, SHG system (C laser) has shown inconsistent changes of power from 10\% to 50\% from the beginning to end of the measurements ranging on measurements that took around 3 hours. 
        In a separate characterization measurement the dye laser (D laser) showed typical power drifts up to $\pm$40\% in 30 minutes.
        Furthermore, during the scans dye laser is expected to have a varying power as the lasing efficiency is frequency dependent.
        
        While no power stability schemes are implemented on either laser, SHG (C laser) frequency is stabilized using a PID loop and showed 6.4~MHz standard deviation over the course of a $\sim6$ hour-long measurement. 
        
        During the measurements, inconsistent drifts on the x-y position of the sample up to $\sim$100~nm/hour are observed.
        One of the reason for these drifts is likely the scanning (galvo) mirrors not being in a temperature stabilized environment (attocube sample stages in the cryostat are expected to have a higher long-term stability).
        Drops in the fluorescence signal of $\sim$10\%/hour are also observed due to the fiber coupling mirrors remaining in a not temperature-stabilized environment.
        
        Finally, while not quantified, z-focus of the sample that is controlled with the z-axis attocube stage is observed to drift over time and required optimization similarly with the x-y position.
        This error is attributed to the heaters not perfectly stabilizing the objective lens in the cryostat to the room temperature (over 100~mK oscillations have been observed).
        It is expected that this problem further contributed to the irregular shifts in the x-y plane as well.
        After the experiments were completed, the heater issue was fixed and consequently the z-axis focus problem was resolved and further stability on the x-y positioning was observed.
       
        During the linewidth broadening measurements, the power of the scanning laser is stabilized using an acousto-optical modulator driven by a microcontroller (Red Pitaya) operating in a PID loop.

\section{Lifetime Extraction for E1}    \label{sec:ap_lifetime}

\begin{figure*} 
            \centering
            \includegraphics[width=\linewidth]{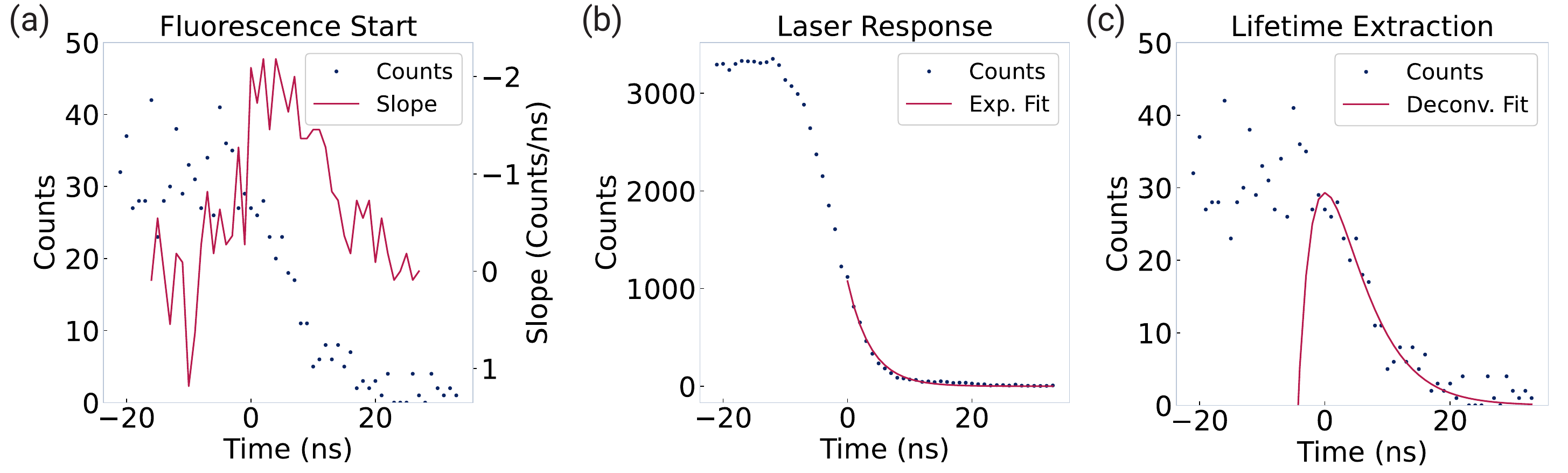}
            \protect\hypertarget{fig:lifetimeL}{}\caption{\label{fig:lifetime} 
                (a) Fluorescence response after exciting the tin-vacancy with a square pulse with a falling edge, overlapped with the stepwise data change (slope). The point of sudden change is selected as time 0. 
                (b) Temporal profile of the excitation laser after being scattered by the sample. The falling edge after time 0 is fit with an exponential function.
                (c) Fluorescence response and the fit function based on the convolution of two exponential decays.}
    \end{figure*}

Due to technical reasons, lifetime measurement for E1 by excitation with a pulse shorter than the lifetime was not possible during the experiments. For having an estimation, we generate a square pulse of 200 ns from a resonant laser transmitted through an acousto-optical modulator and excite the emitter. We then histogram the fluorescence response from the repeated measurements. From these data, we identify the start of the exponential decay following the excitation pulse by calculating the changes between data points (slopes), and determine the threshold for setting the zero time point (Fig.~\ref{fig:lifetime}\hyperlink{fig:lifetimeL}{(a)}). Then, we repeat the measurement without fully filtering the excitation laser and histogram the laser pulse's temporal shape (Fig.~\ref{fig:lifetime}\hyperlink{fig:lifetimeL}{(b)}). We fit the falling curve starting from the previously identified zero point and extract $\tau_{\rm AOM}=3.7(4)$~ns. We attribute the mismatch from the previous measurement for the start of the falling curve before time 0 to a saturated behavior of the fluorescence. Finally, to deconvolve the fall time of the AOM from the lifetime decay, we fit the function to the following equation, which corresponds to the analytical formula for the convolution of two exponential functions (Fig.~\ref{fig:lifetime}\hyperlink{fig:lifetimeL}{(c)}).
\begin{equation}
  e^{-t/ \tau_{\rm lifetime}} \ast  e^{-t/\tau_{\rm AOM}} = \frac{e^{-t/ \tau_{\rm lifetime}}-e^{-t/\tau_{\rm AOM}}}{1/\tau_{\rm AOM}-1/\tau_{\rm lifetime}}
\end{equation}
We set the fitting algorithm such that the maximum of the fitted function always coincides with time 0 and extract $\tau_{\rm lifetime}=5.0(2)$~ns

\section{Temperature Dependent Spectroscopy Analysis Details}    

\begin{figure*}
            \centering
            \includegraphics[width=\linewidth]{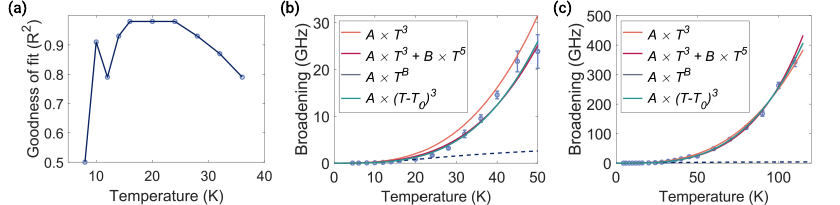}
            \protect\hypertarget{fig:rsquaredL}{}\caption{\label{fig:rsquared} 
                (a) Goodness of fit ($R^2$) value for the fit used for the C transition ($\gamma_+$) with Eq.~\eqref{eq:boltzmann} as the number of data points increases. 24 K is selected for the cutoff threshold into the two-phonon regime, where another function is utilized.
                (b) Different fit functions applied to the data points acquired between 28 and 110 K, plotted over 0-50 K range. It could be observed that the cubic fit does not capture the data very well in the 24-50 K range, while the other functions are qualitatively more successful.
                (c) Different fit functions applied to the data points acquired between 28 and 110 K, same as panel (b), plotted over the full range.}
    \end{figure*}
\subsection{Linewidth analysis}\label{sec:ap_r2}
    In the main text, the linewidth broadening with increasing temperature is investigated to identify the regimes where the broadening is single or two-phonon process dominant.
    These trends give clues about the phononic processes, as in the single phonon regime depolarization events are dominant, and the two-phonon regime is governed by dephasing.
    With this purpose, the data are analyzed to identify a cutoff temperature to distinguish the phononic regimes.

    PLE measurements, at every temperature, are conducted by 10 to 20 frequency scans (in each scan direction) and histogrammed together. 
    Data collected between 4.5 and 32 K temperatures are fitted using a Voigt function, and the data points at 36 and 40 K are fitted using a Lorentzian function, as it yielded lower uncertainties.
    Data above 50 K are acquired with PL emission spectroscopy under non-resonant excitation.
    The C and D transition zero phonon lines are fit using a double Lorentzian function.

    For the C transition, PLE data acquired under 16 K temperature are collected under two configurations: only having the scanning resonant laser and having an additional blue laser in the background as well.
    For the plotted analyses, only the data without the blue laser in the background are used.
    The PLE data between 16 and 40 K are collected with a blue laser ($\sim$100 uW) on the background as the emitter ionized too quickly without collecting enough signal. 
    Since the additional broadenings with the blue illumination in the background under 16 K are measured consistently around $\sim$50 MHz as the linewidth thermally broadened from $\sim$50 to $\sim$400 MHz, the additional diffusion above this temperature is deemed negligible.
    For the D transition, all the data are acquired under additional blue illumination, and due to the reasoning above, spectral diffusion is neglected as the base value for the linewidth is measured to be $\sim$4 GHz.
    An additional test was carried out by fitting the single-phonon regime with a constant offset representing broadening by spectral diffusion $\gamma_{\rm spec}$. These fits, however, yielded $\gamma_{\rm spec}=0$.
    
    All the PLE data have had 32.0(8) MHz subtracted as the natural linewidth to only present the additional broadening.
    At all temperatures, a saturation measurement is carried out, and the employed resonant power is selected as the saturation power. 
    To account for the power broadening, all the data are normalized with a $\sqrt{2}$ factor \cite{OrphalKobin2023PRX}.
    Both directions of the scans are independently analyzed, and their averages are taken into account due to hysteresis effects.
    The linewidths above 40 K are extracted from PL spectroscopy. 
    From the values at each temperature (for both C and D transitions), the measured FWHM extracted at 4.5 K is subtracted to deconvolve the spectrometer resolution jitter. 

    In Fig.~\ref{fig:rsquared}\hyperlink{fig:rsquaredL}{(a)}, the goodness of fit value ($R^2$) is extracted for the C transition while increasing the number of data points from the lower temperature end. The data are fit to Eq.~\eqref{eq:boltzmann} based on Boltzmann distribution provided in the main text.
    A local maximum along 16-24 K with 0.98 value is identified. The value with the largest amount of data points, 24 K, is selected as the threshold for determining the start of the two-phonon process regime.
    Then, to determine the C transition broadening model, four different models are applied:\\
    i) ${A \times T^3}$: Although the theoretically expected function is a cubic trend \cite{Jahnke2015NJP}, as observed in Fig.~\ref{fig:rsquared}\hyperlink{fig:rsquaredL}{(b)} such a polynomial fails to imitate the data between 24 and 50 K. A is extracted to be $2.52 (8) \times 10^{-4}$ GHz/K$^3$. For values above 50 K, a well behaving fit can be observed qualitatively in Fig.~\ref{fig:rsquared}\hyperlink{fig:rsquaredL}{(c)}\\
    ii) ${A \times (T-T_0)^3}$: In Ref. \cite{Goerlitz2020NJP}, to model the experimental data well, a $T_0=10$ K offset is selected for a dataset with only PL spectra with a physical explanation of a thermally lagging sample. Here, such an explanation can not be put forth as the data with the D transition can be reproduced without any offset. It must be noted that D transition and C transition PLE data ($T\leq40$K) were acquired in different temperature scans. $A=2.9(4) \times 10^{-4}$ GHz/K$^3$ and $T_0=4.9(1)$ K are extracted from the fit.\\
    iii) ${ A \times T^B}$: This analysis is carried out for extracting a power dependence of the data set and $A=6(4) \times 10^{-5}$ and $B=3.3(1)$ are extracted. \\
    iv) ${A \times T^3 + B \times T^5}$: This model is selected as a Raman-like two-phonon process can occur with $T^5$ dependence \cite{Jahnke2015NJP}. The extracted coefficients are $A=1.8 (3) \times 10^{-4}$, and $B=8 (6) \times 10^{-9}$.

    For the temperature-dependent spectral shift analysis, a calibration offset is determined from the wavelengths measured independently by PLE and PL spectroscopy at 40 K, and this value is subsequently subtracted from the data acquired above 40 K.
    
\subsection{Spectral shift analysis using the thermal expansion model}\label{sec:ap_shift}
Using the model provided in Ref. \cite{Jahnke2015NJP}, the change in the transition energy $\Delta_{\rm E}$ due to thermal expansion follows the equation
\begin{equation}
\Delta_{\rm E}= - A_{\rm pres} \times B_{\rm dia} \times E(T)
\end{equation}
where the bulk modulus of the diamond $B_{\rm dia}$ is taken to be 443 GPa \cite{McSkimin1972JAP}.
The temperature-dependent expansion coefficient is calculated by the summation
\begin{equation}
    E(T)= \displaystyle\sum_{n=1}^{4} X_i \frac{(\Theta_i/T)^2 e^{(\Theta_i/T)}}{(e^{\Theta_i/T}-1)^2}
\end{equation}
where the $\Theta_i$ and $X_i$ values are taken from Ref.  \cite{Jacobson2019DRM} and $T$ is temperature.
$A_{\rm pres}$ is the hydrostatic pressure shift of the transition energy of the SnV, which is an unknown value and is the first fit parameter separately calculated for both C ($A_{\rm pres,C}=1.1(1)\times10^{-27}$) and D ($A_{\rm pres,D}=3.4(7)\times10^{-27}$) transitions.
For convenient comparison with the phononic model, the energy difference equation is converted to the fit model for the wavelength $\lambda$ using the following equation:
\begin{equation}
    \lambda(T) =\frac{c}{\frac{\Delta_{\rm E(T)}}{h}-\frac{c}{\lambda_0}}
\end{equation}
where $c$ is the speed of light, $ h$ is the Planck constant, and $\lambda_0$ is the wavelength at 0 K, which is the second fit parameter for C ($\lambda_{\rm0, C}=619.061 (2)$ nm) and D ($\lambda_{\rm0, D}=620.111 (1)$ nm) transitions.

\section{E2 Characterisation} \label{sec:ap_emitters}
    The CPT experiments (Sec. \ref{sec:OrbitalCPT}) are conducted on emitter E2.

    \subsection{Autocorrelation} \label{sec:ap_g2}
\begin{figure}
            \centering
            \includegraphics[width=0.33\textwidth]{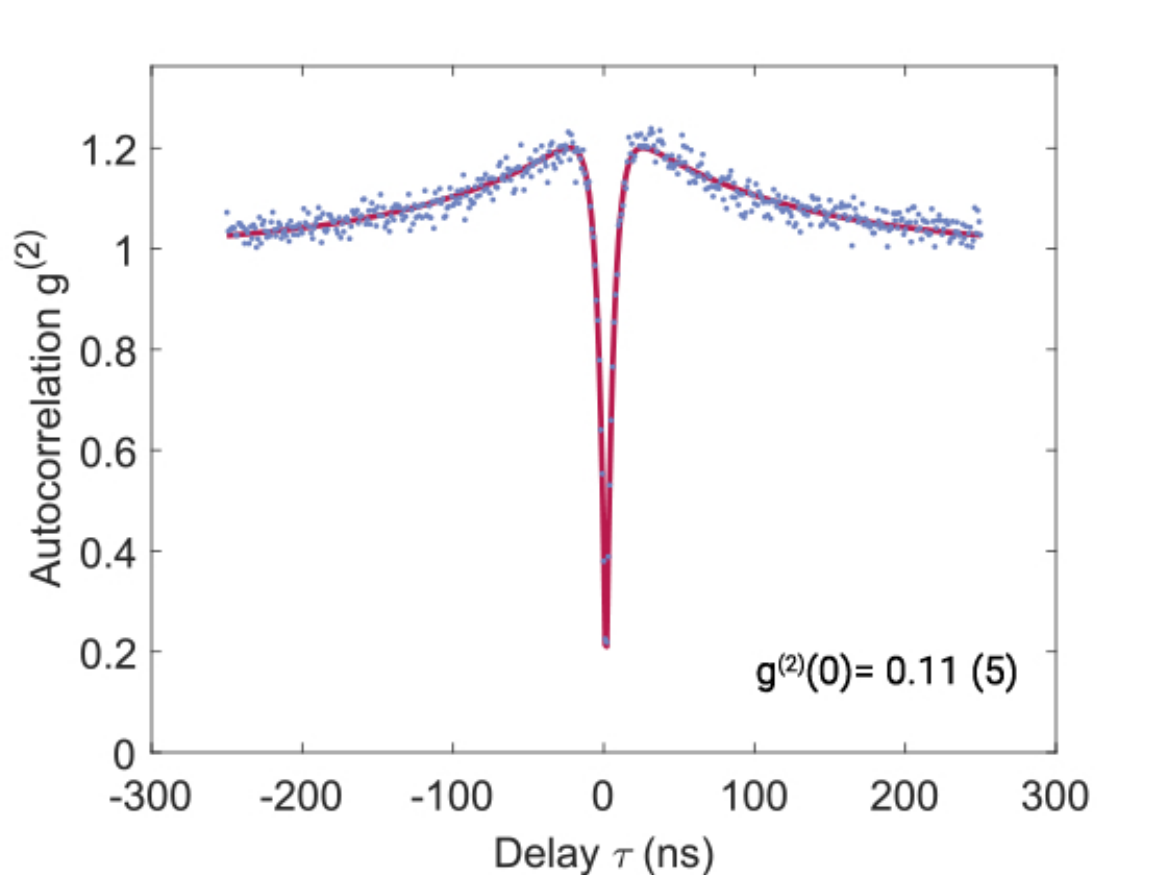}
            \caption{\label{fig:combinedg2} 
                Autocorrelation measurement of E2 under 520~nm, 500~µW non-resonant excitation. The extracted $\rm g^{(2)}(0)$ values confirm that the investigated nanopillar contains a single emitter coupling to its nanostructure.
            }
    \end{figure}

        In order to confirm the investigated emitter is indeed single, an autocorrelation measurement using a Hanbury Brown-Twiss interforemeter configuration is carried out (Fig. \ref{fig:combinedg2}).
        Collected fluorescence under non-resonant excitation is split via a fiber beamsplitter and the coincidences from two ports are correlated.
        Then, the autocorrelation data are fit using the following equation \cite{Kitson1998PRA}:
        \begin{multline}
           {\rm g^{(2)}}(\tau)= 1+p^2\Biggl[c\exp\left(\frac{-|\tau-o|}{\tau_b}\right) \\ -(1+c)\exp\left(\frac{|\tau-o|}{\tau_a}\right) \Biggr]
        \end{multline}
        where $p$, $c$, $\tau_a$ (antibuncing time), $\tau_b$ (bunching time), and $o$ (detector offset) are fitting parameters and $\tau$ is the delay.
    
    \subsection{Spectroscopy} \label{sec:ap_spect}
    
\begin{figure}
        \centering
        \includegraphics[width=0.33\textwidth]{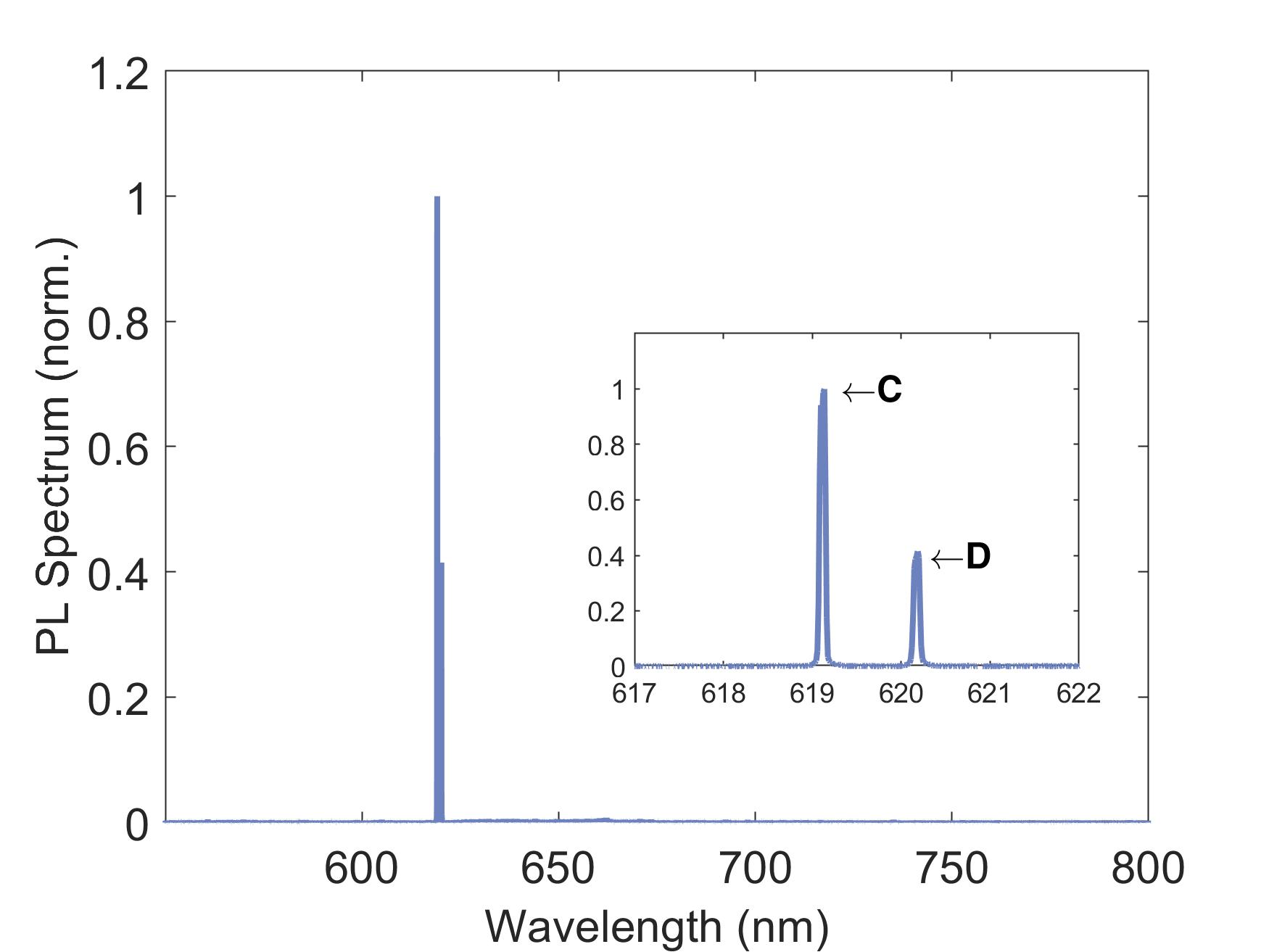}
        \caption{ \label{fig:nihilusSpectrum}
             PL spectrum measurement of the emitter E2 at 4~K under 520~nm green non-resonant excitation. 
             The peak height ratio of 2.4~(1) is extracted for estimating individual spontaneous emission rates in both transitions. 
            }
    \end{figure}

\begin{figure*} 
            \centering
           \includegraphics[width=0.8\textwidth]{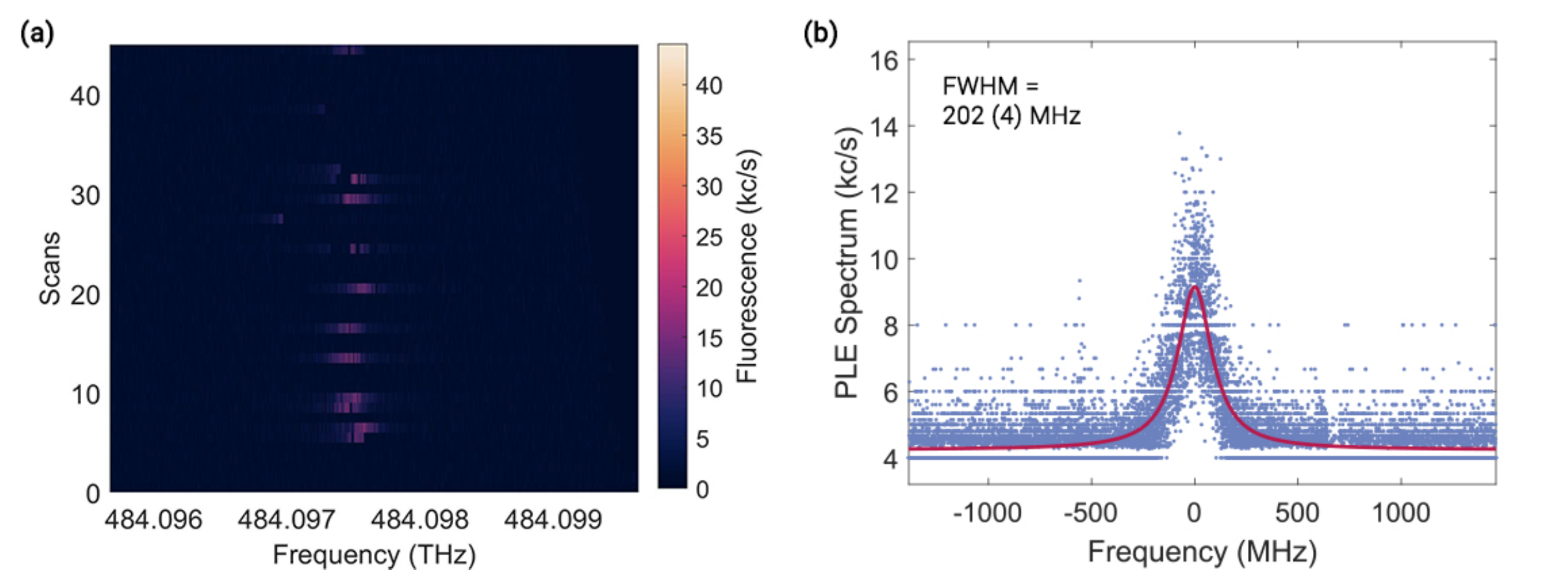}            \caption{\label{fig:nihilusCLinewidth}
                PLE measurement at 4~K to identify the C transition linewidth of the emitter E2. 
                Resonant (C) and blue laser powers are set to 0.5, and 375~nW respectively. 
                Blue laser is pulsed at the beginning of every scan for 10~ms. 
                (a) Frequency scans.
                (b) Histogrammed and fit count rates using a Lorentzian function. The uncertainty is the 95\% confidence interval extracted from the fit.
            }
    \end{figure*}

    In Fig. \ref{fig:nihilusSpectrum} and \ref{fig:nihilusCLinewidth} PL and PLE specta of E2 are provided as characterization measurements of the emitters.
    PL spectrum is utilized to extract information about the branching from the excited state decay for the emitter E2.
    
    \subsection{Lifetime} \label{sec:ap_lifetime}
\begin{figure}
            \centering
            \includegraphics[width=0.33\textwidth]{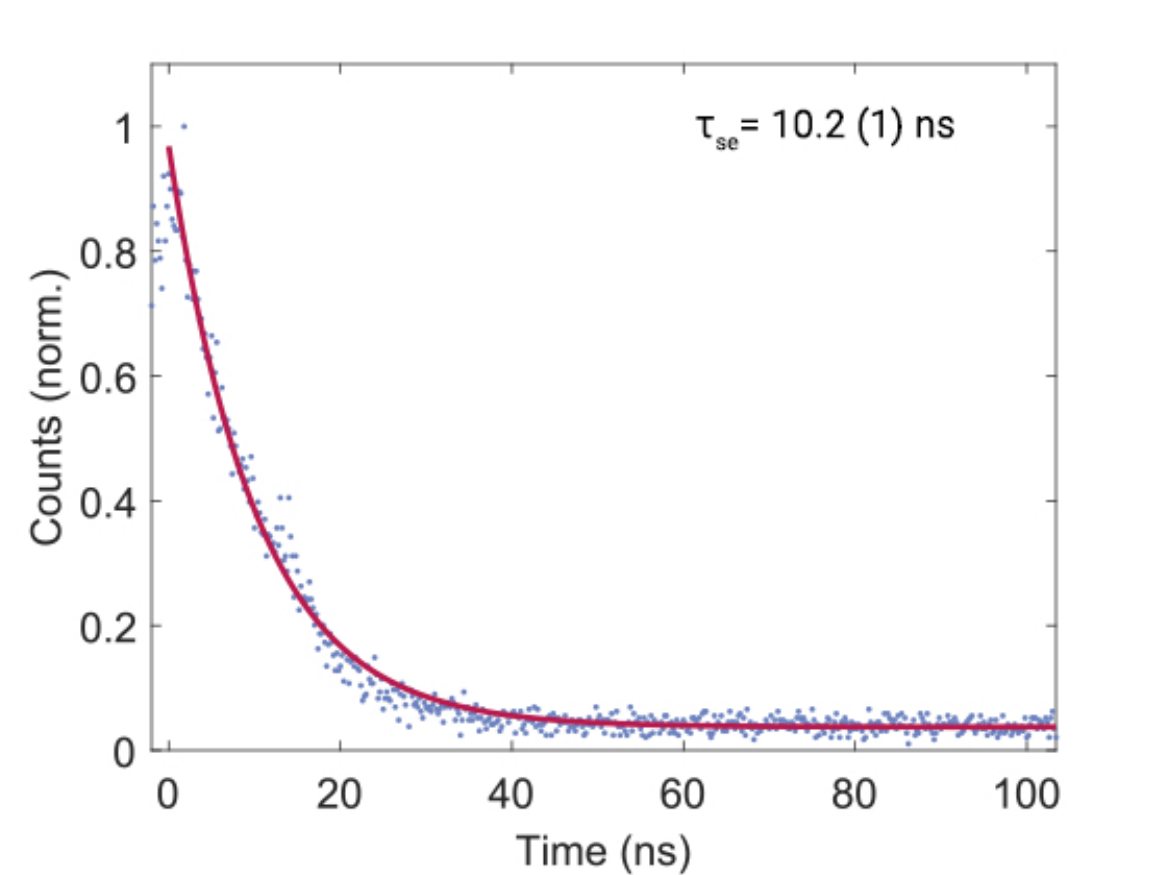}
            \caption{\label{fig:nihilusLifetime}
            Histogrammed fluorescence signal measurements after repetitive resonant 150~fs pulsed excitations.
            Before the excitation, a 500~ns green laser pulse is sent to initialize the charge state of the SnV.
            An exponential decay fit reveals the photonic lifetime of level $\ket{3}$. 
            This value is used to estimate the combined spontaneous emission rate of the C and D transitions and employed in the CPT model.
            The uncertainty is the 95\% confidence interval extracted from the fit.
            }
    \end{figure}

    One of the parameters that are required in CPT modeling is the lifetime $\tau_{\rm se}$ of the excited state.
    A resonant pulsed excitation shorter than the lifetime ($\sim$ 150~fs) is implemented. In Fig. \ref{fig:nihilusLifetime}, the fluorescence response is fit to an exponential $\exp(-\frac{1}{\tau_{\rm se}}t)$ to extract the time $t$ it takes on average for the emitter to spontaneously emit. 
    
    \subsection{Saturation} \label{sec:ap_lifetime}
\begin{figure}
            \centering
            \includegraphics[width=0.33\textwidth]{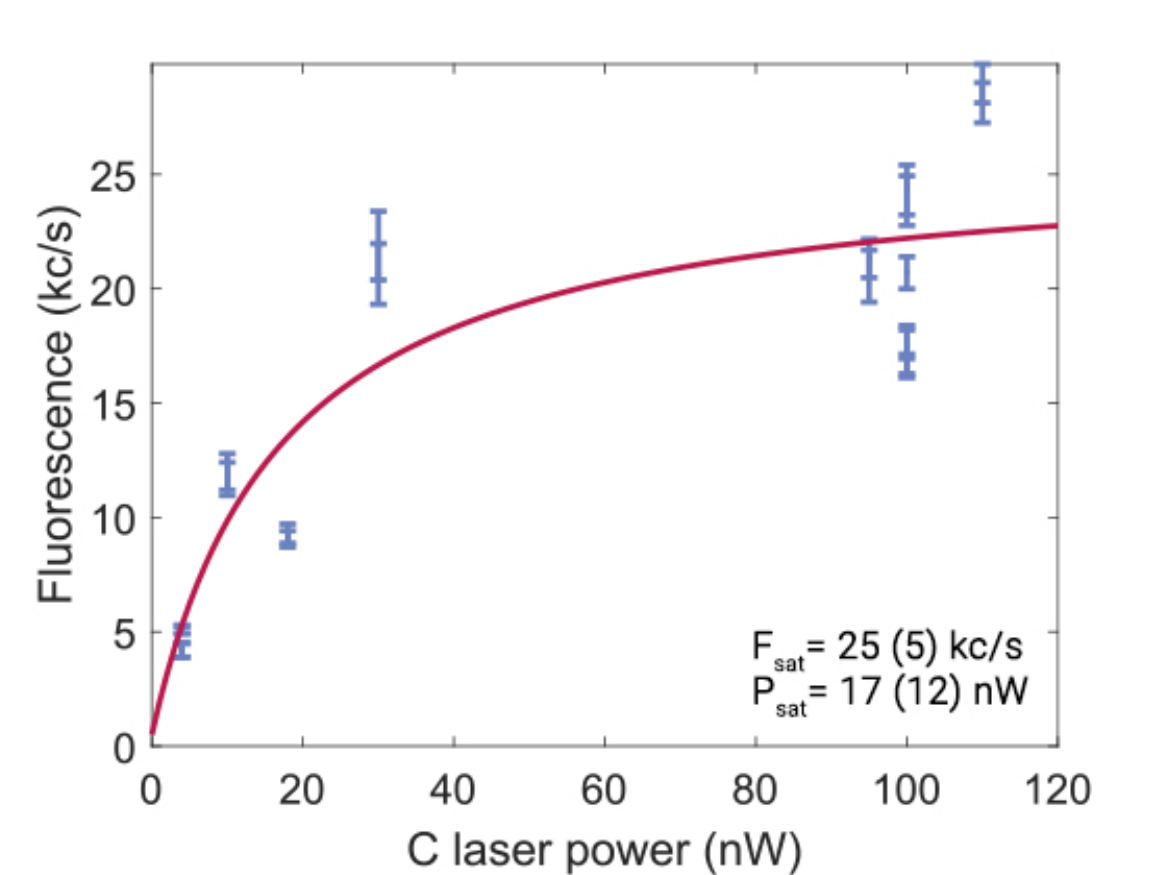}
            \caption{\label{fig:nihilusSat}
                Saturation curve for the emitter E2, using the data extracted from the CPT experiments.
                Error bars are the standard deviation of 100 data points employed to estimate the average countrate.
                The uncertainty is the 95\% confidence interval extracted from the fit.
            }
    \end{figure}

    Translating fluorescence signals to the excited state populations requires an estimation of the saturation intensity.
    With this purpose, in Fig.~\ref{fig:nihilusSat}, the fluorescence signal at the beginning of each CPT measurement is extracted using the average of the first 100 data points (where the D transition detuning is very large and fluorescence from D transition is negligible), and the error bars are estimated from the standard deviation of these data.
    This method had to be implemented because the emitter stopped fluorescing before a typical power-versus-count-rate measurement could be performed.
    
    By pairing the extracted fluorescence counts with the C laser powers at the beginning of each measurement, the saturation fluorescence is estimated using the following equation:
    \begin{equation}
        F(P)=\frac{F_{\rm sat}P}{P_{\rm sat}+P}+0.5,
    \end{equation}
    where $F_{\rm sat}$ (saturation intensity) and $P_{\rm sat}$ (saturation power) are the fitting parameters. $P$ is the power, and 0.5 is selected as the background correction due the dark counts (from APD specifications).
    
\section{CPT Modeling and Analysis}
    \subsection{Lindblad model for CPT dynamics} \label{sec:ap_model}
        In order to model the CPT dip, the evolution of three levels in a lambda scheme is calculated by numerically solving a Lindbladian master equation \cite{Manzano2020AIPAdv}:
        \begin{equation} \label{eq:lindbladian}
            \dot{\rho}=-\frac{i}{\hbar}[H,\rho]+\sum_{ij}\left(S_{ij}\rho S_{ij}^\dagger-\frac{1}{2} \{ S_{ij}^\dagger S_{ij},\, \rho \} \right).
        \end{equation}
        Here, $\rho$ is the density matrix given by $\ketbra{\Psi}{\Psi}$, where $\ket{\Psi_0}=1\ket{1}+0\ket{2}+0\ket{3}$ is the initial state wavefunction. 
        $H$ is the coherent part of the equation and is given by:
        \begin{equation} \label{eq:coherentHam}
            H=\hbar \begin{bmatrix}
                    0 & 0 & \Omega_{\rm C}/2\\
                    0 & \Delta_{\rm C}-\Delta_{\rm D} & \Omega_{\rm D}/2 \\
                    \Omega_{\rm C}/2 & \Omega_{\rm D}/2 & \Delta_{\rm C}
                    \end{bmatrix}
        \end{equation}

        where the matrix includes the Rabi frequencies $\Omega$ on both transitions (C, D) and the detunings $\Delta$ of the electromagnetic field.
        The $S_{ij}$ terms in Eq. \eqref{eq:lindbladian} are operators presenting the incoherent part of the equation and are given by Eq.~\eqref{eq:incoherentOP}, where $i$ and $j$ give the initial and final states respectively:
         \begin{equation} \label{eq:incoherentOP}
            S_{ij}=\sqrt{\gamma_{ij}}\ketbra{j}{i}.
        \end{equation}

        These matrix elements are identified with the following processes; transitions between the ground-state orbital levels:\\
            $ \gamma_{12}= \gamma_+$\\
                $\gamma_{21}=\gamma_-$\\
            decays from the excited state: \\
                   $\gamma_{31}=\gamma_{\rm C}; $\\
                  $ \gamma_{32}=\gamma_{\rm D}; $\\
            dephasing between the ground-state levels: \\
                  $ \gamma_{22}=\gamma_{\rm d} $.
    
        In order to fit the data, Eq.~\eqref{eq:lindbladian} is integrated until the steady state is achieved.
        In this work, 1~$\upmu$s turned out to be sufficient as the integration upper bound.
        This follows the fact that the state evolutions are dominated by $\gamma_-$ that is in the order of few tens of picoseconds.
        
    \label{sec:ap_parameters}
                    \begin{figure*}
                \centering
                \includegraphics[width=\textwidth]{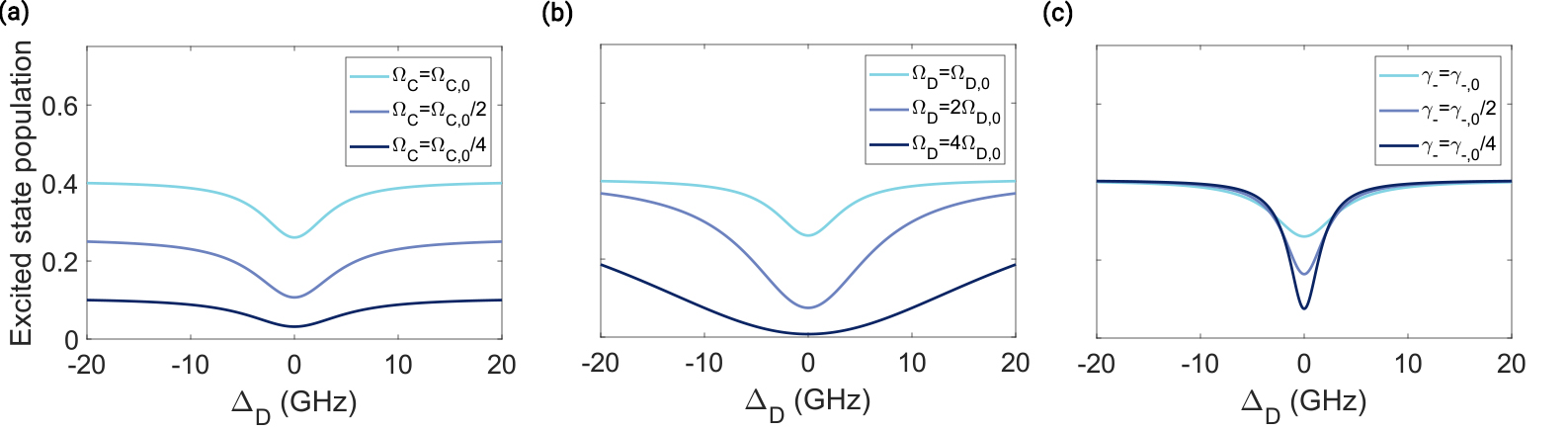}
                \caption{\label{fig:parameterAnalysis}
                    Changes in the CPT dips as fit parameters (with subscripts 0) are varied. Different parameters affect the dip in different ways enabling the implementation of a fitting algorithm with multiple free parameters.
                    (a) CPT dip as the Rabi frequency on the C transition is varied. Change of this parameter mainly determines the excited state population when D is highly detuned.
                    (b) CPT dip as the Rabi frequency on the D transition is varied. Change of this parameter influences both the width and the depth of the dip. 
                    (c) CPT dip as the orbital phononic transition rates (both $\gamma_+$ and $\gamma_-$ via thermalization) are varied. Changing this parameter only influences the depth of the dip.
                }
            \end{figure*}
    
            \subsection{Fit parameter analysis} 
         In Fig.~\ref{fig:parameterAnalysis}, three different parameters (with subscript 0) that are extracted from the fits from measurement number 17 (Appendix \ref{app:fitResults}) are varied. 
         As these three parameters have qualitatively different effects on the dip, one can treat them independently and implement a fitting algorithm.

    In order to estimate a bound for $T_{2}^*$, the orbital dephasing rate $\gamma_{\rm d}$ is increased (from 0) for each fit, until the dip visibility at zero detuning is reduced by 5\%.
    Averaging (standard deviation) over the individual fits, 149~(84)~ps is found to be the upper limit for which the dephasing process can still visibly affect the shape of the dip. 
    This demonstrates that CPT dip is mainly governed by the competition between $\gamma_{\rm -}$ and $\Omega_{\rm D}$ as dephasing is expected to be a much slower process than depolarization at the experimental temperature. 

        \subsection{CPT analysis steps} \label{sec:ap_dataAnalysis}
            \begin{enumerate}[nosep]
                \item The frequencies are scanned between 30 and 200 times (on both directions), depending on the measurement and observed signal.
                \item The frequency scans with the D laser is done via feeding an external voltage signal. 
                As the scan continues, frequency of the laser is recorded through a pick-off. 
                The extracted (voltage, signal) data points are changed into (frequency, signal) by correlating the time stamps of the voltages and frequencies.
                \item All the scans in a measurement are binned according to one of the frequency arrays.
                \item All the bins that have more than 2~kc/s are averaged. 
                The data points with under 2~kc/s are neglected, as the drop in counts indicates measurements where a spectral jump has occurred and the signal is temporarily diminished. 
                \item A pre-fit using a Gaussian function is applied to determine the center of the dip, and the averaged and centered fluorescence data are extracted.
                \item Using the saturation measurement, the fluorescence signal is scaled to a population where the excited state is assumed to be 50\% at saturation.
                \item $\gamma_{\rm C}$+$\gamma_{\rm D}$ is extracted from the lifetime measurement, and their branching (2.4:1) is determined from the PL spectrum.
                \item A second level fit is applied using a numerical black box optimization function in Wolfram Mathmetica (NMinimize), where an error function defined by the sum of $\left|x_{i,{\rm experimental}}-x_{i,{\rm model}}\right|^2$ at each data point, $i$, is minimized by changing the $\Omega_{\rm C}$, $\Omega_{\rm D}$, and $\gamma_{-}$ parameters, where $\gamma_{+}=\gamma_{-}\exp\left[-\frac{\hbar\Delta_{12}}{\kBT}\right]$.
                $\kB$ is the Boltzmann constant, $\Delta_{12}=831{\rm~GHz}$ is the ground state splitting extracted from the resonances identified during a CPT measurement, and $T=3.86{\rm~K}$ is the temperature.
                \item To estimate the fitting sensitivity, each fit parameter is changed on both directions, and the CPT dip population at two-photon resonance is monitored. The maximal value from both directions that changes the dip by more than 5\% is selected as the uncertainty.
            \end{enumerate}

\onecolumngrid   
\clearpage
\subsection{All CPT measurements} \label{sec:ap_allCPT}
            
            \begin{figure} [H]
                \centering
                \includegraphics[width=0.7\linewidth]{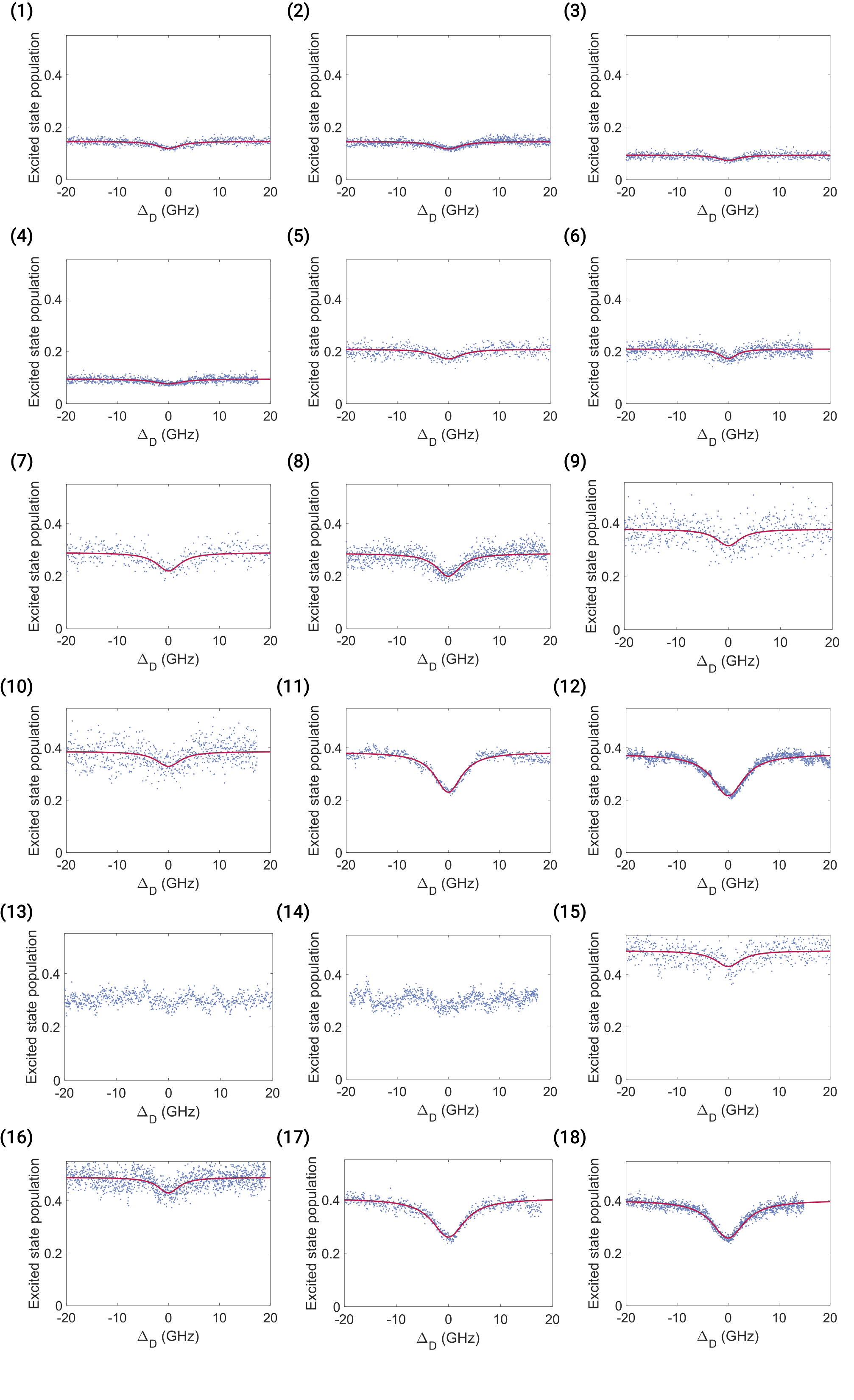}
                \caption{\label{fig:allCPT}
                    All the analyzed CPT data and their fits (solid line) based on the model presented in Appendix \ref{sec:ap_model}. The measurement parameters and the extracted values are provided Appendix \ref{app:fitResults} with the measurement indices.
                }
            \end{figure}
\clearpage
\subsection{Fit results} \label{app:fitResults}
\begin{table}[H]
    \centering
    {\addtolength{\tabcolsep}{3\tabcolsep}
        \begin{tabular}{%
            c@{\hspace{4\tabcolsep}}%
            *{3}{r}@{\hspace{4\tabcolsep}}%
            *{4}{r@{\hspace{.2em}}l}@{\hspace{2\tabcolsep}} %
            r
            }\toprule
            \multirow{2}{*}{Dataset Index} & \multicolumn{3}{@{}l}{Laser power (nW)} & \multicolumn{8}{@{}l}{Extracted fit parameters:} \\
            & C & D & BLUE & \multicolumn{2}{l}{$\Omega_{\rm C}$ (2$\pi$ MHz)} & \multicolumn{2}{l}{$\Omega_{\rm D}$  (2$\pi$ MHz)} & \multicolumn{2}{l}{$T_+$ (ns)} & \multicolumn{2}{l}{$T_-$ (ps)} \\ \hline
            1 & 4 & 1300 & 2500 & 7.1 & (3) & 110 & (14) & 954 & (559) & 31 & (18)  \\ 
            2 & 4 & 1300 & 2500 & 7.1 & (3) & 115 & (15) & 915 & (547) & 30 & (18) \\ 
            3 & 4 & 1500 & 2500 & 5.3 & (2) & 104 & (11) & 1156 & (726) & 38 & (24) \\
            4 & 4 & 1500 & 2500 & 5.3 & (2) & 112 & (14) & 846 & (499) & 27 & (16) \\ 
            5 & 10 & 1800 & 2500 & 9.4 & (4) & 124 & (17) & 855 & (489) & 28 & (16) \\ 
            6 & 10 & 1800 & 2500 & 9.4 & (4) & 99 & (14) & 1326 & (762) & 43 & (25) \\ 
            7 & 18 & 1800 & 2500 & 13.0 & (6) & 159 & (15) & 987 & (651) & 32 & (21)  \\ 
            8 & 18 & 1800 & 2500 & 12.8 & (6) & 177 & (13) & 1036 & (727) & 34 & (24) \\ 
            9 & 30 & 2500 & 2500 & 19.3 & (14) & 164 & (24) & 966 & (552) & 31 & (18) \\ 
            10 & 30 & 2500 & 2500 & 20.4 & (17) & 164 & (26) & 921 & (493) & 30 & (16) \\ 
            11 & 95 & 6000 & 4000 & 20.1 & (10) & 272 & (13) & 994 & (741) & 32 & (24) \\ 
            12 & 95 & 6000 & 4000 & 19.3 & (9) & 296 & (14) & 869 & (652) & 28 & (21)) \\ 
            13 & 100 & 160 & 5800 & \multicolumn{2}{l}{$-$} & \multicolumn{2}{l}{$-$} & \multicolumn{2}{c}{$-$} & \multicolumn{2}{c}{$-$} \\
            14 & 100 & 160 & 5800 & \multicolumn{2}{l}{$-$} & \multicolumn{2}{l}{$-$} & \multicolumn{2}{c}{$-$} & \multicolumn{2}{c}{$-$} \\
            15 & 110 & 1500 & 2500 & 82 & (20) & 417 & (65) & 862 & (477) & 28 & (15) \\ 
            16 & 110 & 1500 & 2500 & 74 & (18) & 353 & (58) & 1048 & (569) & 34 & (18) \\ 
            17 & 100 & 6300 & 2500 & 22.9 & (13) & 306 & (16) & 807 & (595) & 26 & (19) \\ 
            18 & 100 & 6300 & 2500 & 22.4 & (13) & 307 & (16) & 785 & (579) & 25 & (19) \\  \hline
            Average (Standard dev.) &  &  &  &  &  & & & 958 & (138) & 31 & (5)  \\\bottomrule
        \end{tabular}
        \caption{\label{tab:}
            Tabulated summary of the coherent population trapping experiments. 
            Subsequent measurements with the same laser parameters are two different scan directions (odd (even): low to high (high to low) frequencies).
            The quoted laser powers are measured at the beginning of the experiments and subjected to drifts during the scans.
            Combined with instability of the sample position, varying Rabi frequencies ($\Omega$) to power relations are observed.
            Provided uncertainties are the 5\% parameter sensitivities at the two-photon resonance (see Appendix \ref{sec:ap_dataAnalysis}).
            Resulting orbital level lifetimes ($T_\pm$) are averaged in the end.
        }
        \addtolength{\tabcolsep}{-3\tabcolsep}
    }
\end{table}

   \twocolumngrid            
\section{Phononic broadening of the E2's D transition at 4 K} \label{sec: ap_Dbroadening}

        \begin{figure}
            \centering           \includegraphics[width=0.33\textwidth]{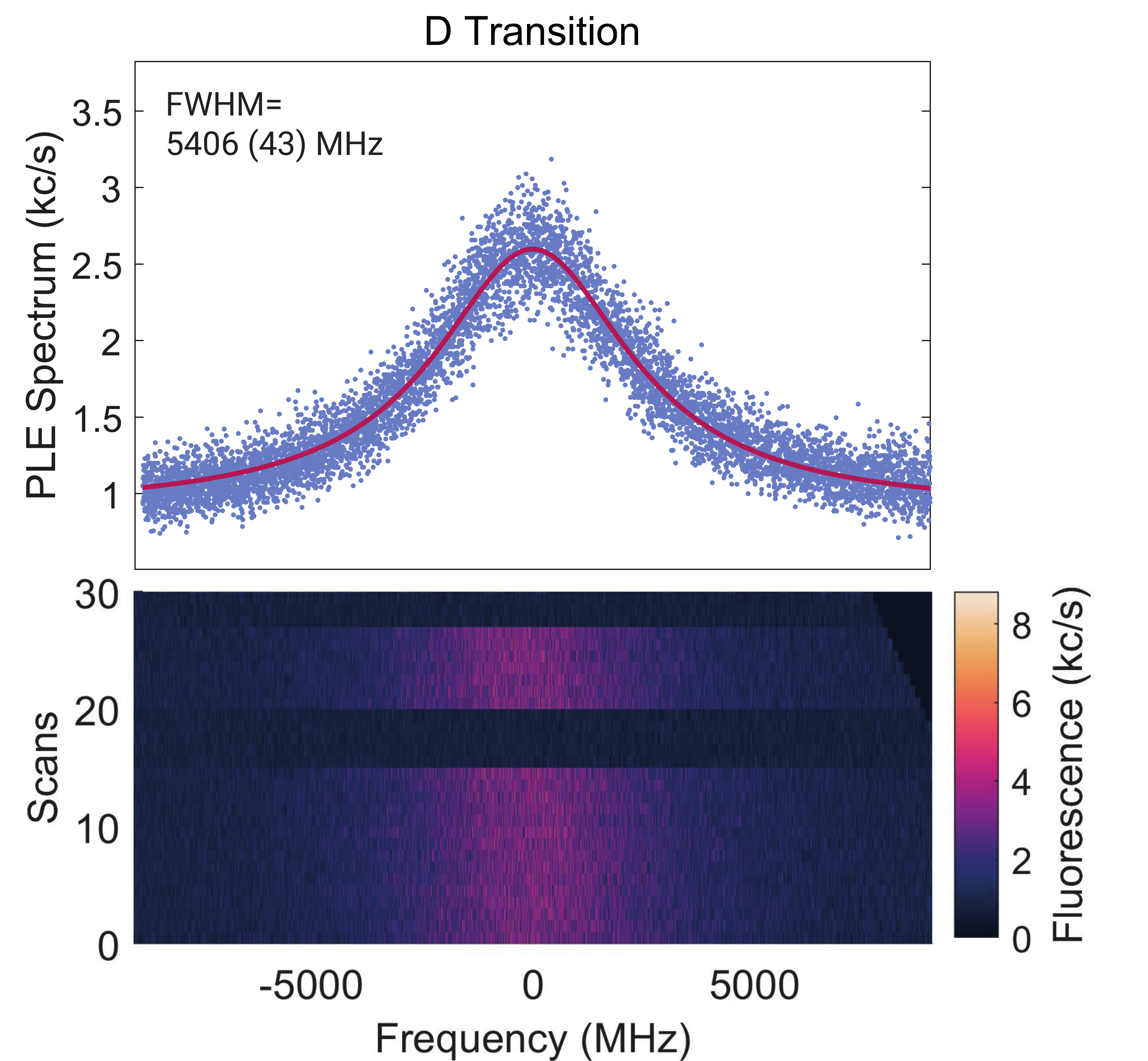}  
          \caption{\label{fig:nihilusDLinewidth}
                 PLE measurement at 4~K to identify the D transition linewidth of the emitter E2. 
                 Resonant laser power is set to 800~nW. 
                 A green laser with 1 µW power is continuously on throughout the measurement for charge state stabilization. 
                 A Lorentzian function is used to fit the data for extracting the FWHM and uncertainty. 
                 A spectral jump in the D transition is observed during the dark periods in the line scans.
                 Since the population is very low in D due to thermalization, a high power is used to extract a signal.
                TOP: Histogrammed and fitted count rates,
                BOTTOM: Frequency scans.
                }
    \end{figure}

Due to the fast thermal relaxation from the upper branch of the ground state, the population in level $\ket{2}$ is extremely low (less than 0.1\%, using the extracted values from above). 
As such, locating and resonantly exciting the D transition can be challenging. 
For example, Fig.~\ref{fig:nihilusDLinewidth} shows a linewidth measurement for the D transition where a resonant power of 800~nW, 8000 times higher than $\sim$100~pW for the C transition, was required in order to detect a very low signal.
This measurement resulted in a linewidth of $\Gamma_{\rm D}={5.406~(43){\rm~GHz}}$. These thermalized populations influence the CPT dips as well, and a minimum $\sim$500~nW of power was needed to observe a dip.

 Using the analysis provided in Ref. \cite{Wang2024PRL}, the photonic components of the C and D transition linewidths $\Gamma_{\rm hom}$ are calculated from the lifetime $\tau_{\rm se}$ presented in Fig.~\ref{fig:nihilusLifetime}:

\begin{equation} \label{eq:dhom}
    \Gamma_{\rm hom}=1/(2\pi\tau_{\rm se})
\end{equation}

The ratio of saturation powers of E1 at 4.5 K data point from Fig.~\ref{fig:TDepCLine} is used to estimate D transition saturation power $P_{\rm sat,D}$ of E2 from the C transition saturation power $P_{\rm sat}$ presented in Fig.~\ref{fig:nihilusSat}:

\begin{equation}
    P_{\rm sat,D}=P_{\rm sat}\frac{P_{\rm sat,D,E1}}{P_{\rm sat,C,E1}}
\end{equation}

From these values, the power broadening components for the C (D) transition $\Gamma_{\rm C(D),pow}$ are calculated:

\begin{equation}
\Gamma_{\rm C,pow}=\Gamma_{\rm hom}\sqrt{1+P_{\rm C}/P_{\rm sat}}-\Gamma_{\rm hom}
\end{equation}
\begin{equation}
\Gamma_{\rm D,pow}=\Gamma_{\rm hom}\sqrt{1+P_{\rm D}/P_{\rm sat,D}}-\Gamma_{\rm hom}
\end{equation}

From the measured C transition linewidth $\Gamma_{\rm C}$ presented in Fig.~\ref{fig:nihilusCLinewidth}, the spectral diffusion component for both transitions $\Gamma_{\rm diff}$ is estimated by
\begin{equation}
\Gamma_{\rm diff}=\Gamma_{\rm C}-\Gamma_{\rm hom}-\Gamma_{\rm C,pow}
\end{equation}

This method, however, assumes the D transition couples to the external charge noise exactly the same as the C transition \cite{Santis2021PRL,Aghaeimeibodi2021PRApp}, therefore should only be treated as an estimation.

Finally, by subtracting all the remaining components, we extract the phononic broadening component $\Gamma_{\rm D,phon}$ of the D transition linewidth:
\begin{equation}
\Gamma_{\rm D,phon}=\Gamma_{\rm D}-\Gamma_{\rm hom}-\Gamma_{\rm D,pow}-\Gamma_{\rm diff}
\end{equation}

$T_{\rm -}$ depends on the $\Gamma_{\rm D,phon}$ by a Fourier transform relation
\begin{equation}
T_{\rm -,D}=1/(2\pi\Gamma_{\rm D,phon})
\end{equation}

$T_{\rm -,D}=30.6 (2)$ ps agrees very well with the $T_{\rm -}=31 (5)$ ps previously calculated from the orbital CPT measurements and thus confirms the value of the ultrafast phononic processes.

%

\end{document}